\documentclass[11pt,a4paper,oneside]{article}
\usepackage[top=3cm, bottom=3cm, left=2cm, right=2cm]{geometry}
\usepackage{amsthm,amsmath,amsfonts,amssymb}
\usepackage[]{natbib}
\usepackage{graphicx}
\usepackage[T2A,T1]{fontenc}
\usepackage[utf8]{inputenc}
\usepackage[russian,english]{babel}
\usepackage[dvipsnames]{xcolor}
\usepackage{amsmath}
\usepackage{mathrsfs}
\usepackage{mathtools}
\usepackage{bbm}
\usepackage{verbatim}
\usepackage{algorithm}
\usepackage{dsfont}
\usepackage{bbm}
\RequirePackage[colorlinks,citecolor=blue,urlcolor=blue,backref=page,backref=page, linkcolor=blue]{hyperref}
\usepackage{multirow}
\usepackage[noend]{algpseudocode}
\usepackage{dsfont}
\usepackage{bm}
\usepackage{algorithmicx}
\usepackage{siunitx}
\usepackage{adjustbox}
\def\simind{\stackrel{\mbox{\footnotesize ind}}{\sim}}
\def\simiid{\stackrel{\mbox{\footnotesize iid}}{\sim}}
\def\CRP{\textsc{CRP}}
\def\d{\mbox{d}}
\def\Pr{\text{Pr}}
\usepackage{textcomp}
\usepackage{subfigure}
\newtheorem{prop}{Proposition}
\usepackage{amsthm,amsmath,amssymb,mathrsfs,dsfont,mathtools}
\usepackage{bm}
\usepackage{graphicx}
\usepackage[colorinlistoftodos]{todonotes}
\usepackage{hyperref}
\usepackage{booktabs}
\usepackage[all]{nowidow}
\usepackage{setspace}
\usepackage{algorithm}
\usepackage[noend]{algpseudocode}
\usepackage{tikz} 
\usepackage[sf]{titlesec}
\usepackage{sectsty}
\allsectionsfont{\centering \normalfont\scshape} % Make all sections centered, the default font and small caps
\usepackage{lipsum}
\usepackage{adjustbox}
\usepackage{lineno} 
\theoremstyle{plain}

\newtheorem{corollary}{Corollary}
\theoremstyle{definition}

\theoremstyle{remark}

\usepackage{authblk}

\title{Local Level Dynamic Random Partition Models\\ for Changepoint Detection}
\date{}
\author[1]{Alice Giampino}
\author[1]{Bernardo Nipoti}
\author[2]{Marina Vannucci}
\author[3]{Michele Guindani} 

\affil[1]{Department of Economics, Management and Statistics, University of Milano-Bicocca, Milan, Italy}
\affil[2]{Department of Statistics, Rice University, TX, U.S.A.}
\affil[3]{Department of Biostatistics, University of California Los Angeles, CA, U.S.A.}

\begin{document}

\maketitle

\begin{abstract}
Motivated by an increasing demand for models that can effectively describe features of complex multivariate time series, e.g. from sensor data in biomechanics, motion analysis, and sports science, we introduce a novel state-space modeling framework where the state equation encodes the evolution of latent partitions of the data over time. Building on the principles of dynamic linear models, our approach develops a random partition model capable of linking data partitions to previous ones over time, using a straightforward Markov structure that accounts for temporal persistence and facilitates changepoint detection. The selection of changepoints involves multiple dependent decisions, and we address this time-dependence by adopting a non-marginal false discovery rate control. This leads to a simple decision rule that ensures more stringent control of the false discovery rate compared to approaches that do not consider dependence. The method is efficiently implemented using a Gibbs sampling algorithm, leading to a straightforward approach compared to existing methods for dependent random partition models. Additionally, we show that the proposed method can be adapted to handle multiview clustering scenarios. Simulation studies and the analysis of a human gesture phase dataset collected through specific sensing technologies show the effectiveness of the method in dynamically clustering multivariate time series and detecting changepoints.
\end{abstract}

\section{Introduction}\label{sec:intro}

Recently, there has been an increased demand for models that can effectively describe features of complex multivariate time-series data. This surge in interest is particularly prominent in fields such as biomechanics, motion analysis, human-computer interaction, and sports science.  Here, one of the goals is often to break down continuous human motion or activity into distinct phases and states.
The insights derived from such analyses can then be used to improve sports performances and for a deeper study of the complexity of human biomechanics, for example, to understand gait cycles, gesture phases, athletic movements, and even cognitive states. In the analysis of human gesture data, information is typically gathered through various sensing technologies, such as motion capture systems, accelerometers, or videos. These technologies yield rich datasets that capture the temporal patterns of human movements. Understanding how the movements change, interact, and unfold over time is essential for capturing the underlying processes driving human motion and behavior. This temporal evolution holds key information about transitions between movement phases, variations in performance, and subtle cues related to coordination and intent, all of which are critical for developing accurate and interpretable models. Such data are commonly collected as multivariate time series, indexed by discrete time points 
$t=1, \ldots,T$, where each time point corresponds to a multivariate observation. We let $\bm{Y}_{1:T}=\{\bm{Y}_1,\ldots, \bm{Y}_T\}$ denote a multivariate  time series observed on $n$ units over $T$ time points, where each $\bm{Y}_t$ is a $n$-dimensional vector, i.e., $ \bm{Y}_t = (Y_{1,t}, Y_{2,t}, \ldots, Y_{n, t})$. To illustrate, consider our application in Section \ref{sec:gesturedata}, where we examine scalar velocity data obtained from
acceleration sensor units placed on the hands and wrists of a subject. Dynamic linear models (DLMs) are commonly employed for the analysis of time-series data due to their flexibility and adaptability in handling diverse situations \citep[][]{petris2009dynamic}.  They define a class of state-space models and are characterized by a system of two equations: an observation equation, which describes the observed data as a linear combination of latent state variables with noise; and a state equation that describes how latent states evolve over time, thereby tracking the underlying dynamics of the system. We introduce our contribution by referring to a simple yet fundamental DLM,  the \emph{local level model}, which describes the observed data as composed of a level component plus random noise, 
\begin{equation}
Y_{i,t} = \beta_{i, t} + \varepsilon_{i,t},
\label{eq:local_level_obs}
\end{equation}
where $\varepsilon_{i,t} \simiid \text{N}(0,\tau^2)$, with $\text{N}(\mu, \sigma^2)$ denoting the Normal distribution with mean $\mu$ and variance $\sigma^2$. The vector $\bm{\beta}_{t} = ( \beta_{1,t}, \ldots, \beta_{n,t})$ represents the underlying level or trend of the time series at time $t$. In a typical local level model, the evolution of \( \bm \beta_t \) over time is modeled as a random walk, i.e., the level at time \( t \) is predicted to be the same as the level at time \( t-1 \), plus some random noise.  If the variance of the random noise is small, this assumption implies smooth processes over time. Despite its simplicity, the local level model illustrates the fundamental characteristics of many time-series models, and will serve as a basic example throughout.

\begin{figure}[t!]
    \centering
    \includegraphics[clip, trim=0.7cm 0.7cm 0.5cm 0.5cm,width=5cm]{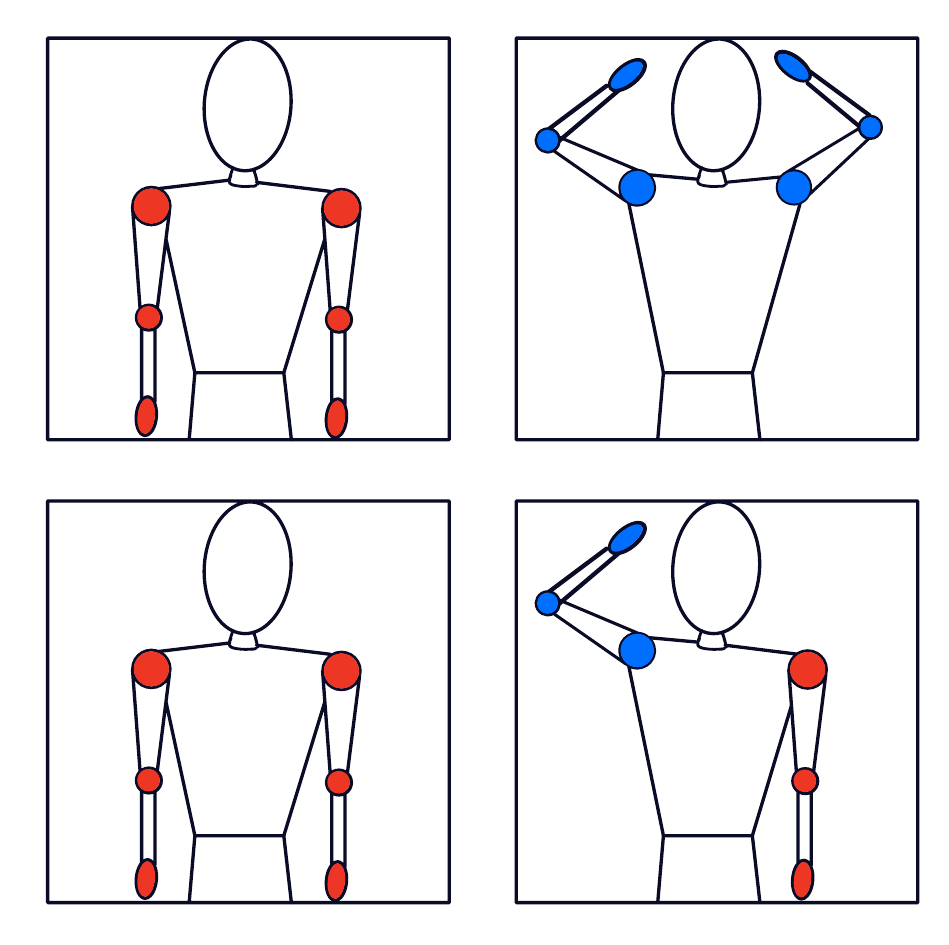}
    \caption{Example of sensors' clusters at two-time points (left, right). Colors identify the clusters of sensors, defined based on the measurement of their scalar velocity. Top: the arms are clustered together in both images, despite a change in the cluster label.  Bottom: the scalar velocity measured across sensors defines two distinct partitions, shown in the left and right images.}
    \label{fig:ex_sensors}
\end{figure}

One of the main goals of our biometric data application is to study how measurements from different units cluster or group over time,  as well as how these clusters evolve during an activity, to understand how different body parts cooperate during different stages of a gesture or movement. Furthermore, the detection of changepoints in the clustering of body parts between gesture stages can reveal important insights into the motion and inform gesture detection algorithms. In the Bayesian setting, Bayesian nonparametric (BNP) models are a popular choice for clustering dynamic behavior over time \citep{quintana2022dependent}. BNP models do not require the upfront specification of the number of clusters; instead, they allow for posterior inference on cluster allocations directly from the data. Existing BNP approaches for clustering time-series data vary in terms of motivation, application, and how time-dependence is introduced. 
For example, some approaches build on the stick-breaking representation of the Dirichlet Process \citep{ferguson1973bayesian, sethuraman1994constructive}. In this context, \citet{antoniano2016nonparametric} have developed a stationary Markov model where both the transition and stationary densities are nonparametric infinite mixture models. \citet{nieto2014bayesian} have clustered temporal data while considering several features typical of time-series data (e.g., trends and seasonality). BNP autoregressive model are discussed in \citet{kalli2018bayesian}, \citet{de2023bayesian}, \citet{deyoreo2018modeling}, \citet{pavani2025bayesian}, and \citet{beraha2023childhood},  among others. Alternatively, other authors have explored  generalizations of the  P\'olya urn scheme of \citet{Bla73}, see, e.g. \citet{caron2007}, \citet{caron2017generalized}  and \citet{Cassese2019}. While %in the motivating example shown in Figure \ref{fig:ex_sensors} 
our goal is to capture changes in the clustering structure of the data over time (see the motivating example shown in Figure \ref{fig:ex_sensors}), where a \textit{changepoint} corresponds to a shift in the partition of the sensors across successive time points, all the methods mentioned above identify clusters based on the values of associated parameters, e.g. the latent variables $\beta_{i,t}$ in the illustrative local level model \eqref{eq:local_level_obs}. 
These models can detect changepoints if, for instance, two data points assume a common value at times $t$ and $t+1$ but that value changes significantly between time points.
This occurs because model-based clustering with Bayesian nonparametric models essentially relies on specifying a mixture model that depends on a discrete random mixing measure. The probabilistic distribution over random partitions is essentially a by-product of this mixture setup; in essence, clusters in a mixture model are identified based on the values of the process' atoms. We illustrate this point in Figure \ref{fig:ex_sensors} in the case of clustering of sensors, where the grouping may not depend on the intensity of the values but rather on the dependence in their movements.

An alternative approach involves directly modeling the sequence of random partitions. In this context, \citet{page2022dependent} introduced a dependent random partition model \citep{hartigan1990partition, barry1992product} that incorporates auxiliary variables that determine whether a unit, for which the cluster allocation at time $t-1$ has been determined, should be considered for possible cluster reallocation at time $t$. As a result, changes in partitions over time are obtained as a by-product of individual clustering assignments. 
\citet{paganin2023informed} proposed a model for a sequence of random partitions over time, considering individual allocation probabilities for each unit rather than the whole partition at each time point. \citet{cremaschi2023change} introduced a Bayesian changepoint model for spatiotemporal data, where partition changes in areal units over time are governed by a spatial association parameter. Their model employs a random partition prior that incorporates spatial features, promoting the co-clustering of nearby areal units based on proximity.
\citet{quinlan2022joint} presented a method that aims to correlate partitions with the detection of changepoints in multivariate time series in the context of contagion phenomena. However, following \citet{martinez2014nonparametric}, they specify only a single unique partition of contiguous clusters for each time series. The focus of their model is on multivariate changepoint detection rather than data partitioning, with changepoints occurring at different time points for each time series. Recently, \citet{corradin2024model} proposed a Bayesian model for clustering time series, where two observations are assigned to the same group if they exhibit structural changes in their behavior at the same time. Their approach assumes that units within the same cluster share changepoints across all time points, with no possibility of shifts of units in different clusters at different time points.

In this paper, we  propose viewing the evolution of partitions over time as latent partition-based states within a state-space model. Our approach develops a random partition model (RPM) capable of linking the partition of data points to previous partitions over time, using a straightforward Markov structure to model these changes. The approach builds upon the principles of DLMs but extends them to incorporate latent state equations now operating within the partition context, with time evolution defined by the partitions themselves. Therefore, while in the analysis of time-series data, it is common to identify sudden changes in the observed values of a stochastic process as a changepoint, in this paper, we refer to a changepoint as a change in a latent partition of units. More specifically,  we employ a Markov-dependent structure, where the partition at time $t$, denoted as $\pi_t$, is modeled conditionally on the partition at time $t - 1$, accounting for temporal persistence and enabling changepoint detection. The selection of changepoints at each time point $t$ is driven by a mixture that chooses between one of two partitions. 
At any given time, the chosen partition can either remain the same as the one at the previous time point or be sampled from a flexible and general random partition model. In the presence of a changepoint, the partition at time $t$ becomes independent of the partition at time $t-1$. We demonstrate that, based on our construction, the random partitions \(\pi_t\), for \(t = 1, \ldots, T\), are marginally identically distributed, with the marginal distribution that can be specified so to ensure tractability. Unlike the existing Bayesian nonparametric models cited above, our clustering approach does not directly rely on the values of parameters, such as the $\beta_{i,t}$'s in equation \eqref{eq:local_level_obs}. Instead, we directly treat dependent random partition allocations as latent structures that drive the dynamics of the observations. In contrast to the dependent random partition model proposed by \citet{page2022dependent}, we jointly consider all the units when identifying changepoints, taking into account the multivariable nature of the dependent partitions. 

For posterior inference, our approach relies on the efficiency of a Gibbs sampling algorithm, offering a straightforward implementation compared to existing methods for dependent random partition models. The  selection of changepoints involves dependent multiple decisions, and we propose accounting for this time-dependence by adapting the decision-theoretic approach of \citet{chandra2019non} in defining the error and non-error terms of the decision loss function. The resulting procedure establishes a simple decision rule that ensures more stringent control of the false discovery rate compared to an approach that does not consider dependence. 

Recent methods have been proposed for multiview clustering that allow for dependence between several experimental conditions or features. For example, \citet{franzolini2023conditional} employ a conditional partially exchangeable model to induce dependence between clustering configurations of the same subjects across different features. Similarly, \citet{dombowsky2025product} develop a nonparametric prior that generates dependent random distributions by centering a Dirichlet process on a random product measure, with a single parameter controlling the dependence. 
While our approach is designed to model the evolution of partitions over time, as in the motivating application shown in Figure  \ref{fig:ex_sensors}, it may be adapted to multiview clustering scenarios.
In such cases, an analogous latent-state representation of the partitions can be employed to build a hierarchical model that effectively captures the dependence between partitions, describing changes across different groups or experimental conditions. Additionally, we show that the dependence between the partitions generated by the hierarchical model for multiview clustering in \citet{dombowsky2025product} coincides with that one of a reparametrization of our prior specification. 
This is not, however, the focus of this work and we note that moving beyond a two-view setting introduces additional computational challenges, which cannot be addressed by a straightforward adaptation of the %two-view 
algorithm designed for time-series data, and thus require careful treatment to preserve both efficiency and robustness.
Finally, we should note that
while constructing the prior distribution for the partition bears some resemblance to the use of spike-and-slab  priors for variable selection \citep{tadesse2021handbook}, dealing with partitions introduces additional complexity in the motivation, modeling and computation.

The article is organized as follows: in Section \ref{sec:model}, we present our model and discuss its main properties,  including the alternative multiview formulation discussed in Section \ref{sec:hierarcLDDP}. Section \ref{sec:post} addresses posterior inference, including computational methods and decision theory-based changepoint detection. In Section \ref{sec:sim} we describe two simulation studies that highlight the key aspects of our model. In Section \ref{sec:application}, we present an application to the analysis of human gesture data, and illustrate its performance in two-view clustering using data from a collaborative perinatal project. Finally, concluding remarks and future directions are provided in Section \ref{sec:fine}. Proofs and additional details on computations and illustrations are available in the Supplementary Material.

\section{Local level dynamic partition model}\label{sec:model}

We describe the key features of the dynamic partition model we propose, a local level dynamic partition model (LLDPM) taking model \eqref{eq:local_level_obs} as reference. More specifically, we assume that for each unit $i$, with $i=1, \ldots, n$, the observations are generated from some general likelihood (observation equation) as
\begin{eqnarray}\label{eq:likelihood}
Y_{i, t} \mid \beta_{i, t} \stackrel{\text { ind }}{\sim}{p}\left(y_{i,t} \mid \beta_{i,t}\right),
\end{eqnarray}
for $t=1, \ldots T$.  Throughout this work, we consider a Gaussian kernel with mean \(\beta_{i,t}\) and variance \(\tau^2\). However, it is possible to extend this construction to other kernels. The dynamics of the latent state equation are characterized in terms of time-varying partitions of the $n$ units over time. In order to describe such dynamics,  we introduce a dependent RPM that models temporal dependence in terms of sequences of partitions by considering an auxiliary variable determining whether the partition at time \(t-1\) will be revised for possible reallocation of the units at time \(t\), for \(t = 1, \ldots, T\). More specifically, 
we introduce a binary \emph{changepoint} auxiliary variable $\gamma_t \in \{0,1\}$, to detect changes in the partitions of the $n$ units from time $t-1$ to time $t$. We use $|\cdot|$ to indicate the cardinality of a set, and thus denote the number of blocks composing a partition $\pi_t$ of $n$ units as  $\left|\pi_t\right|$. Further, we let $\pi_t=\{C_{1,t}, \ldots, C_{\left|\pi_t\right|, t}\}$, where, for each $j=1,\ldots,\left|\pi_t\right|$, the set $C_{j, t}$ consists of the indices corresponding to the statistical units assigned to the $j$-th cluster according to $\pi_t$. Thus, given a partition $\pi_{t-1}$ at time $t-1$, 
we assume a \emph{partition-based state equation} characterized as a mixture over two partition models, corresponding to the case of conditionally independent and identical random partitions at subsequent time points. Namely,
\begin{equation}
\begin{aligned}
\pi_{t} \mid \bm{\pi}_{1:(t-1)}, \bm{\gamma}_{2:t} &\sim\left(1-\gamma_{t}\right) \delta_{\pi_{t-1}} (\pi_t) +\gamma_{t} \, p^*(\pi_t),\quad t=2, \ldots, T \\
\pi_1&\sim p^*(\pi_1),
\label{eq:partition_state_equation}
\end{aligned}
\end{equation}
where $\bm{\pi}_{1:(t-1)}=(\pi_1,\ldots,\pi_{t-1})$ is the vector of previously recorded partitions,  $\bm{\gamma}_{2:t}=(\gamma_2,\ldots,\gamma_{t})$ indicates the vector of changepoints, and $p^*(\cdot)$ is a distribution over the space of partitions of $n$ units, henceforth referred to as the \emph{base partition distribution}. In addition, we assume the changepoints $\gamma_t$ in \eqref{eq:partition_state_equation} are distributed according to
\begin{equation}\label{eq:bern}
\gamma_t\stackrel{\text{ind}}{\sim} \operatorname{Bern}\left(\eta_{t}\right),\quad t=2, \ldots, T.
\end{equation}
The probability of a changepoint, $\eta_t$, can then be seen as a dependence parameter, with the extreme case $\eta_t=0$ leading to almost surely identical partitions at time $t-1$ and $t$. In the remaining of the work, we will call \emph{partition state model} (PSM) the model for the random partitions $\bm{\pi}_{1:T}$ defined through \eqref{eq:partition_state_equation} and \eqref{eq:bern}.   Alternative partition-based state equations, and thus partition state models, can potentially be defined,  to describe more complex forms of temporal dependence. See Section \ref{sec:fine} for more discussion. The LLDPM is thus obtained by combining the PSM with \eqref{eq:likelihood}, where, as standard in Bayesian nonparametric models, we assume that at each time point $t=1,\ldots,T$, the values of the parameters $\beta_{i,t}$, with $i=1,\ldots,n$, coincide within a cluster, although the specific values could differ at different times. If we let 
$\bm{\beta}^\ast_t = (\beta_{1,t}^*,\ldots,\beta_{|\pi_t|,t}^*)$ denote the vector of $|\pi_t|$ distinct values in $\bm{\beta}_t$, as determined by $\pi_t$, such assumption can be formalized as
   $\bm{\beta}^\ast_t 
   \sim \prod_{j=1}^ {\left|\pi_t\right|} P_0\left(\beta_{j, t}^*\right)$,
for some base probability distribution $P_0\left(\cdot\right)$.

\subsection{Prior properties of the PSM}

\noindent We assume that the base partition distribution $p^*(\cdot)$ is the one implied by the class of Gibbs-type priors \citep{de2013gibbs}. Distributions over partitions are conveniently described through the exchangeable partition probability function (EPPF), a simple way to define probabilistic partition models based on the number and sizes of blocks, independently of the object labels. \citet{de2013gibbs} show that exchangeable product partition models with probability of each partition depending only on the cardinality of each cluster coincide with the family of Gibbs-type priors. A Gibbs-type EPPF has the following form,
\begin{equation}\label{eq:eppf_gibbs}
\mathbb{P}_\theta\left[\pi_t=\{C_{1,t}, \ldots, C_{|\pi_t|, t}\}\right]=V_{n, |\pi_t|} \, \prod_{j=1}^{ |\pi_t|} \frac{\Gamma\left(|C_{j, t}|-\sigma\right)}{\Gamma(1-\sigma)},
\end{equation}
with $-\infty \leq \sigma<1$, and where, for any $n\geq1$ and any $k=1,\ldots,n$, the set of non-negative weights $V_{n,k}$ satisfies the following recursive equation 
    $V_{n,k}=(n-\sigma k) V_{n+1,k}+V_{n+1,k+1}$, and is such that $V_{1,1}=1$.
Notable elements of the class of Gibbs-type priors are the Dirichlet and the Pitman--Yor processes \citep{ferguson1973bayesian, pitman1997two}, mixtures of symmetric
Dirichlet distributions \citep{gnedin2005exchangeable}, the normalized inverse Gaussian processes \citep{lijoi2005hierarchical} and the normalized generalized gamma process \citep{lijoi2007controlling}. Throughout the paper, we use the Chinese restaurant process \citep[CRP,][]{pitman2006combinatorial} and its generalization, the two-parameter CRP \citep[2-CRP,][]{pitman2006combinatorial}, as running examples for the base partition distribution $p^*(\cdot)$. The CRP, denoted $p_{\text{CRP}}(\cdot)$, is parametrized by a concentration parameter $\theta > 0$ and corresponds to distribution on the partitions induced by the Dirichlet process. The corresponding EPPF is readily obtained by setting $\sigma=0$  and  $V_{n, |\pi_t|}=\theta^{|\pi_t|}/(\theta)_n$ in \eqref{eq:eppf_gibbs}, 
where \((\theta)_n=\theta(\theta+1)\cdots(\theta+n-1)\) denotes the ascending factorial. The 2-CRP, denoted $p_{\sigma}(\cdot)$, introduces an additional discount parameter $\sigma \in [0,1)$, with $\theta > -\sigma$, and coincides with the distribution on the partitions associated with the Pitman--Yor process \citep{Per92,pitman1997two}. The corresponding EPPF is obtained by setting $V_{n,|\pi_t|}=\prod_{j=1}^{|\pi_t|-1}(\theta+j\sigma)/(\theta+1)_{n-1}$ in \eqref{eq:eppf_gibbs}. For each specification within the class of Gibbs-type priors, the weights $V_{n,k}$ can be computed recursively or estimated via Monte Carlo \citep[see][]{Arb17}.
 
\begin{figure}[h!]
\centering
\subfigure{\includegraphics[clip,trim=0.6cm 1.4cm 3cm 0.4cm,height=3.25cm, page=6]{our_dependence01_quadrato1.pdf}}\hspace{0.8cm}
\subfigure{\includegraphics[clip,trim=0.6cm 1.4cm 3cm 0.4cm,height=3.25cm, page=5]{our_dependence01_quadrato1.pdf}}\hspace{0.8cm}
\subfigure{\includegraphics[clip,trim=0.6cm 1.4cm 0cm 0.4cm,height=3.25cm, page=4]{our_dependence01_quadrato1.pdf}}\\[-2pt]
\subfigure{\includegraphics[clip,trim=0.6cm 1.4cm 3cm 0.4cm,height=3.25cm, page=3]{our_dependence01_quadrato1.pdf}}\hspace{0.8cm}
\subfigure{\includegraphics[clip,trim=0.6cm 1.4cm 3cm 0.4cm,height=3.25cm, page=2]{our_dependence01_quadrato1.pdf}}\hspace{0.8cm}
\subfigure{\includegraphics[clip,trim=0.6cm 1.4cm 0cm 0.4cm,height=3.25cm, page=1]{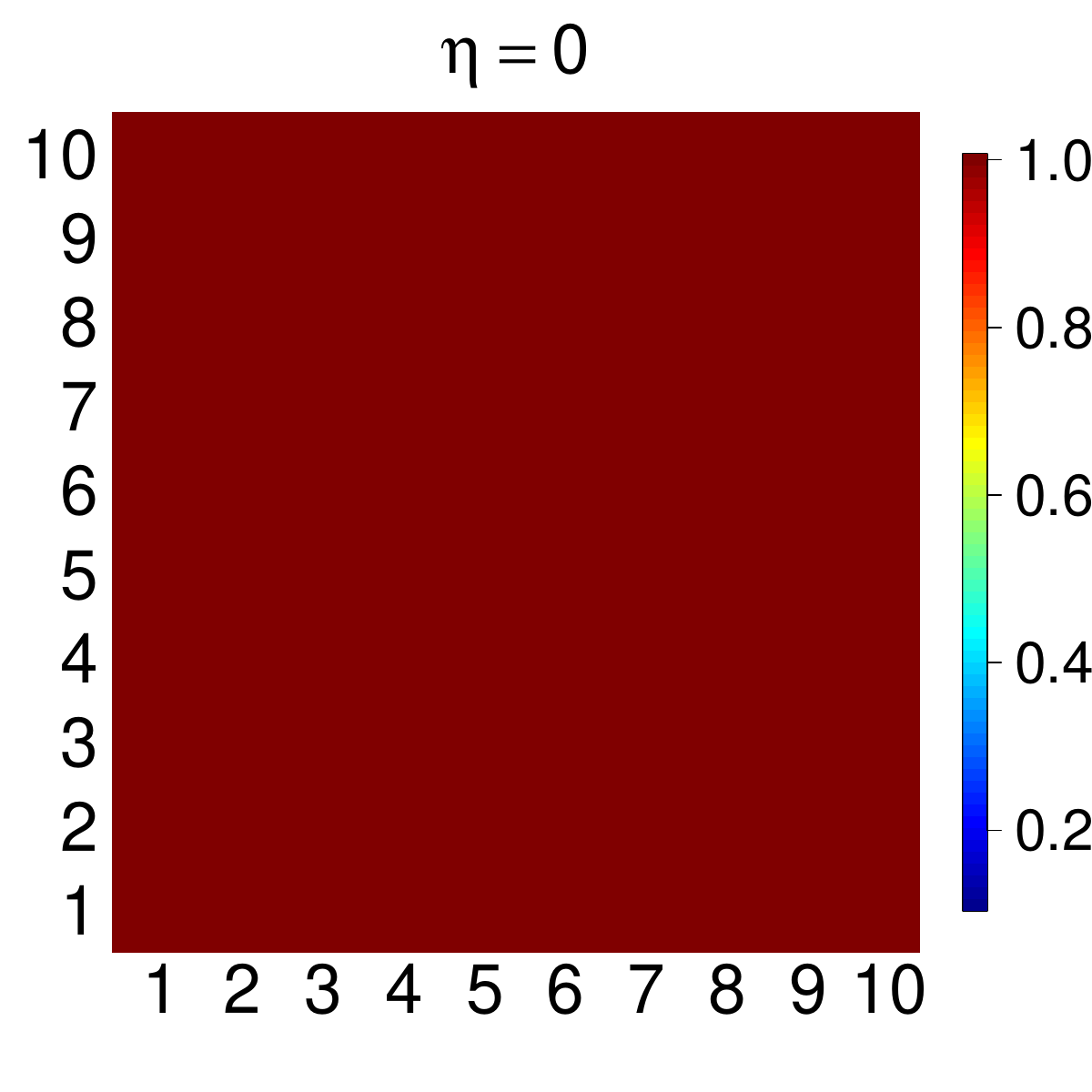}}\\
\caption{Average lagged ARI for the pairwise comparison of $T=10$ random partitions $\bm{\pi}_{1:T}$ modeled according to the proposed PSM with $p^*(\cdot)=p_{\text{CRP}}(\cdot)$. From top-left panel to bottom-right panel, $\eta$ ranges in $\{0.95,0.75,0.5,0.25,0.05,0\}$. For each matrix, the pixel in position $(i,j)$ refers to the comparison of $\pi_{i}$ and $\pi_{j}$. For each value of $\eta$, values of the lagged ARI are averaged over a sample of 10,000 partitions. The temporal dependence increases as the temporal dependence parameter $\eta$ decreases.
}\label{fig:temp_dep}
\end{figure}

We study the temporal dependence of random partitions implied by the proposed PSM. We begin by observing that, once the changepoints $\bm{\gamma}_{2:T}$ are marginalized out, the distribution of $\bm{\pi}_{1:T}$ implied by the PSM is given by
\begin{align*}
\bm{\pi}_{1:T}&\sim p^*(\pi_1) \, \prod_{t=2}^T \left[(1-\eta_t)\, \delta_{\pi_{t-1}}(\pi_t)+\eta_t\, p^*(\pi_t)\right].
\end{align*} 
Figure  \ref{fig:temp_dep} shows the average Adjusted Rand Index (ARI) \citep{rand1971objective} between time-lagged partitions, similarly to Figure 2 in \citet{page2022dependent}, assuming $T=10$, for several values of the temporal dependence parameter, $\eta_t=\eta\in\{0,0.05,0.25,0.5,0.75,0.95\}$, $t=2,\ldots,T$. For each value of $\eta$ and for each pair of times $t_1,t_2\in\{1,\ldots,T\}$, the average lagged ARI, comparing $\pi_{t_1}$ and $\pi_{t_2}$, is computed over 10,000 partitions simulated from the proposed PSM with $p^*(\cdot)=p_{\text{CRP}}(\cdot)$. Figure \ref{fig:temp_dep} shows increasing temporal dependence of the lagged partitions as 
$\eta$ decreases. The dependence across  partitions at different times can be investigated more formally by computing the expected Rand Index (ERI). For two random partitions $\pi_{t_1}$ and $\pi_{t_2}$, the Rand Index is a random variable defined as  $R(\pi_{t_1}, \pi_{t_2}) = (A_{t_1,t_2}+B_{t_1,t_2})/\binom{n}{2}$, where $A_{t_1,t_2}$ 
is the number of pairs of observations that are in the same cluster in both partitions, and $B_{t_1,t_2}$ the number of pairs that are in different clusters in both partitions. The corresponding ERI, $\varphi_{t_1,t_2}$, 
is then defined as 
$$ 
\varphi_{t_1,t_2}=\mathbb{E}[R(\pi_{t_1}, \pi_{t_2})] = \sum_{\pi_{t_1}, \pi_{t_2} \in \mathscr{P}} R(\pi_{t_1}, \pi_{t_2})\, p(\pi_{t_1}, \pi_{t_2}),
$$ 
where $\mathscr{P}$ indicates the space of all possible partitions of $n$ elements, and $p(\pi_{t_1},\pi_{t_2})$ stands for the joint distribution of the random partitions $\pi_{t_1}$ and $\pi_{t_2}$. We obtain the ERI for two random partitions whose joint distribution is defined through the proposed PSM and the base partition distribution $p^*(\cdot)$ implied by a Gibbs-type prior. We first focus on partitions at subsequent time points, that is $t_2=t_1+1$, and then consider the more general case $t_1<t_2$. 
\begin{prop}\label{prop:ERI}
Suppose $t_2=t_1+1$ and \( \eta_{t_1} = \eta_{t_2} = \eta \). Let \( p^\ast(\cdot) \) be the distribution over partitions implied by a Gibbs-type prior. Then the ERI for two random partitions $\pi_{t_1}$ and $\pi_{t_2}$ modeled according to a PSM is given by
\begin{equation}\label{eq:ERI}
\varphi_{t_1,t_2} = 1 - 2V_{2,2}(1 - V_{2,2})\eta.
\end{equation}
\end{prop}
\noindent The ERI is thus a decreasing function of the dependence parameter value $\eta$. Moreover, as \(\eta \to 0\), the ERI approaches one, its maximum value, indicating that the partitions are identical. Conversely, as \(\eta \to 1\), the ERI converges to $1-2V_{2,2}(1-V_{2,2})$, which is the ERI of two independent and identically distributed partitions with distribution implied by a Gibbs-type prior. The expression in \eqref{eq:ERI} can be rewritten as a convex linear combination of these two limiting cases, that is
$\varphi_{t_1,t_2}=(1-\eta) + [1-2V_{2,2}(1-V_{2,2})] \eta$.

When $p^*(\cdot)=p_{\text{CRP}}(\cdot)$ or $p^*(\cdot)=p_{\sigma}(\cdot)$, the ERI for two random partitions at subsequent times modeled according to the proposed PSM is obtained by substituting the appropriate specification of \( V_{2,2} \) in Proposition \ref{prop:ERI}, which leads to the following corollary.
\begin{corollary}
Suppose $t_2=t_1+1$ and \( \eta_{t_1} = \eta_{t_2} = \eta \). 
Then the ERI for two random partitions $\pi_{t_1}$ and $\pi_{t_2}$ modeled according to a PSM 
is given by
\begin{align}\label{eq:ERI_DP}
\varphi_{t_1,t_2} &= 
1 - \frac{2 \theta}{(\theta+1)^2}\eta& \text{if }p^*(\cdot)=p_{\CRP}(\cdot),\\\label{eq:ERI_PY}
\varphi_{t_1,t_2} &= 1- \ \frac{2( \theta + \sigma) (1-\sigma)}{(\theta+1)^2}\eta&\text{if }p^*(\cdot)=p_{\sigma}(\cdot).
\end{align}
\end{corollary}
\noindent The expression for the ERI in \eqref{eq:ERI_DP} coincides, up to a reparametrization of the dependence parameter $\eta$, with an analogous result derived in \citet{dombowsky2025product}. As expected, the ERI in \eqref{eq:ERI_DP} is recovered by setting $\sigma=0$ in \eqref{eq:ERI_PY}. Interestingly, the ERI approaches one both when \(\theta\to -\sigma\) and when \(\theta\to \infty\). This behavior is not surprising as, when \(\theta\) approaches $-\sigma$, the distributions of both random partitions tend to assign all their mass to the partition with only one block. Similarly, when $\theta$ increases, the two distributions accumulate mass at the partition composed of $n$ blocks of size one. In both cases, the two random partitions tend to take the same configuration, regardless of the value taken by the dependence parameter $\eta$. Also in this case, \eqref{eq:ERI_DP} and \eqref{eq:ERI_PY} can be written as convex linear combinations of the ERI of two identical random partitions, that is one, and the ERI of two independent and identically distributed random partitions, that is $\varphi_{t_1,t_2}=(\theta^2+1)/(\theta+1)^2$ if $p^*(\cdot)=p_{\CRP}(\cdot)$ and $\varphi_{t_1,t_2}=(\theta^2+1)/(\theta+1)^2+2\sigma(\theta+\sigma-1)/(\theta+1)^2$ if $p^*(\cdot)=p_{\sigma}(\cdot)$. 

The next proposition describes how the dependence between two lagged random partitions modeled through the proposed PSM changes as a function of the lag $t_2-t_1$.
\begin{prop}\label{prop:ERIt1t2}
    Suppose $t_{1}<t_{2}$ and let $\eta_{t_{1}}=\eta_{t_{2}}=\eta$. Let \( p^\ast(\cdot) \) be the distribution over partitions implied by a Gibbs-type prior. Then the ERI for two random partitions $\pi_{t_1}$ and $\pi_{t_2}$ modeled according to a PSM 
    is given by
    $$
    \varphi_{t_1,t_2} = 1-2 V_{2,2}(1-V_{2,2})\left[1-(1-\eta)^{t_2-t_1}\right].
    $$
\end{prop}
\noindent The ERI is thus a decreasing function of the time lag \(t_2 - t_1\). Furthermore, regardless of the value of the dependence parameter $\eta$, the limit of \(\varphi_{t_1, t_2}\) as \((t_2 - t_1) \to \infty\) tends towards the ERI of two independent random partitions with identical distribution implied by a Gibbs-type prior. The next corollary specializes the previous result to the cases $p^*(\cdot)$ being a CRP or a 2-CRP.
\begin{corollary}
    Suppose $t_{1}<t_{2}$ and let $\eta_{t_{1}}=\eta_{t_{2}}=\eta$. 
    Then the ERI for two random partitions $\pi_{t_1}$ and $\pi_{t_2}$ modeled according to a PSM is given by 
    \begin{align}\label{eq:ERI_DPt1t2}
    \varphi_{t_1,t_2} &= 1-\frac{2\theta}{(\theta+1)^2}[1-(1-\eta)^{t_2-t_1}]& \text{if }p^*(\cdot)=p_{\CRP}(\cdot),\\\label{eq:ERI_PYt1t2}
    \varphi_{t_1,t_2} &= 1-\frac{2(\theta+\sigma)(1-\sigma)}{(\theta+1)^2}[1-(1-\eta)^{t_2-t_1}]&\text{if }p^*(\cdot)=p_{\sigma}(\cdot).
     \end{align}
\end{corollary}

We next explore the marginal distribution of the components of $\bm{\pi}_{1:T}$, under the assumption of a PSM for $\bm{\pi}_{1:T}$, with generic base partition distribution $p^*(\cdot)$. 
We prove that at each time $t=1,\ldots,T$, the marginal distribution of the random partition $\pi_t$ is the same as that of the base partition distribution $p^*(\cdot)$. 
Thus, although the PSM  explicitly models the evolution of the random partitions $\bm{\pi}_{1:T}$ over time, the components of $\bm{\pi}_{1:T}$ are marginally identically distributed according to $p^{*}(\cdot)$. Importantly, this result holds regardless of the specific distribution $p^{*}(\cdot)$, which therefore could be specified also outside the class of Gibbs-type priors considered here.
\begin{prop}\label{prop:PSM_marginal}
Let 
$\bm{\pi}_{1:T}$ be modeled according to a PSM. 
Then, for every $t=1,\ldots,T$, the marginal distribution of the random partition $\pi_t$ coincides with the base partition distribution $p^*(\cdot)$.
\end{prop}

\subsection{ Multiview clustering representation of the LLDPM}\label{sec:hierarcLDDP}

In many real-world applications, data can be collected from various sources or represented using different feature sets, creating multiple views of the same underlying entities, with each view defined on its own support.
In this section, we present an alternative representation of our model that establishes a hierarchical structure in the dependence of partitions, deviating from the assumption of temporally dependent partitions where each partition depends on the previous one. This hierarchical representation effectively describes changes in the partition across different groups or experimental conditions, offering a flexible and interpretable approach to modeling and understanding partition relationships. Such representation exemplifies a multiview clustering model, bearing close connections to the partition introduced by \citet{dombowsky2025product} and related to the multiview representation in \citet{franzolini2023conditional}. 
As noted below \eqref{eq:ERI_DP}, the same expected dependence between partitions induced by the hierarchical model for multiview clustering in \citet{dombowsky2025product} can be achieved by our model by reparametrizing the dependence parameter $\eta$. In their prior specification, dependence is introduced through the concentration parameter of a Dirichlet process, which determines the tendency of data points to form distinct groups. The main difference between our model and theirs lies in the way the dependence is induced. While their models specify a dependence structure on the atoms within a hierarchical model on a product space, our model introduces dependence directly at the level of partitions, providing a more streamlined approach to modeling partition relationships. 

Without loss of generality, we begin by examining the case $T=2$,  
where \(T\) can now be interpreted as the number of different views, groups or experimental conditions, depending on the context of the application. Thus, we focus on the joint distribution of $\pi_1$ and $\pi_2$, as implied by the PSM, which is 
\begin{align}\label{eq:joint_pi2}
\bm{\pi}_{1:2}&\sim p^*(\pi_1)\left[(1-\eta)\delta_{\pi_{1}}(\pi_2)+\eta\,p^*(\pi_{2})\right],
\end{align}
where $\eta$ is used to denote the dependence parameter. This distribution can be obtained as the marginal distribution of a hierarchical random partition model. To see this, we consider a third random partition distributed according to the base partition distribution, namely 
\begin{equation}\label{eq:alternative_model_1}
\tilde\pi\sim p^*(\tilde\pi).
\end{equation}
In this alternative representation, $\tilde\pi$ serves as a common parent partition for both $\pi_1$ and $\pi_2$, by assuming the following generative model for  $\pi_1$ and $\pi_2$,  
\begin{align}
\begin{split}\label{eq:alternative_model_2}
        \pi_{t}\mid \tilde\pi, \tilde{\gamma}_1,\tilde{\gamma}_2 &\simind (1-   \tilde{\gamma}_t)\, \delta_{\tilde\pi} (\pi_t)+\tilde{\gamma}_t\, p^*(\pi_t),\qquad  t=1,2,
\end{split}
\end{align}
where $\tilde{\gamma}_t \simiid \text{Bern}(\tilde{\eta})$, $t=1,2$, controls the whether $\pi_t$  coincides with the parent partition $\tilde\pi$. 
Higher values of $\tilde\eta$ increase the chances of $\pi_t$ being different from $\tilde\pi$, allowing for more flexibility and individual variations in the partition structure. Conversely, a lower value of $\tilde\eta$ makes it more likely that $\pi_t$ will follow the same partition structure as $\tilde\pi$, enforcing a stronger dependence between the two. Note that,  in model \eqref{eq:alternative_model_1}--\eqref{eq:alternative_model_2}, the distribution of $\pi_2$ is independent of that of $\pi_1$,  given the partition $\tilde\pi$. The joint distribution of the random partitions $\pi_1$ and $\pi_2$, the common parent partition $\tilde{\pi}$, and the dependence variables $(\tilde{\gamma}_1,\tilde{\gamma}_2)$, here denoted by $p(\bm{\pi}_{1:2},\tilde\pi,\tilde{\gamma}_1,\tilde{\gamma}_2)$, can thus be written as
\begin{align*}
    p(\bm{\pi}_{1:2},\tilde\pi,\tilde{\gamma}_1,\tilde{\gamma}_2)&\sim 
    p^*(\tilde\pi)\, \left\{\prod_{t=1}^2 [(1-  \tilde\gamma_t)\delta_{\tilde\pi}(\pi_t)+\tilde\gamma_t\, p^*(\pi_t)]\,%\times
    \tilde\eta^{\tilde\gamma_t}(1-\tilde\eta)^{1-\tilde\gamma_t}\right\}.
\end{align*}
The distribution of $\pi_1$ and $\pi_2$ is obtained by marginalizing the last expression with respect to $\tilde\pi$ and $(\tilde\gamma_1,\tilde\gamma_2)$, 
which gives \eqref{eq:joint_pi2}, provided that $\tilde{\eta}=1-\sqrt{1-\eta}$. See Section \ref{sm:multiview} in the Supplementary Material.
Hence, the local level partition model's joint distribution on the partitions can be viewed as a special case of a hierarchical partition model, when considered marginally. This representation highlights the connection between the time-varying formulation of the PSM, as specified through \eqref{eq:partition_state_equation}-\eqref{eq:bern}, and an exchangeable model on the partitions, under the assumptions of identical marginal distributions of the partitions, $p^*(\pi_t)$, and equal Bernoulli probabilities $\tilde\eta$ across groups. This hierarchical representation can be extended to accommodate $T \geq 2$ groups, with one parent random partition driving the distribution of $T$ identically distributed random partitions. Moreover, the assumption of a common Bernoulli probability $\tilde{\eta}$ across groups can be relaxed to accommodate for group-specific parameters $\tilde{\eta}_t$, with $t=1,\ldots,T$. Expressing the local level partition model as a hierarchical model provides an alternative viewpoint to appreciate its flexibility in capturing both similarities and dissimilarities among the partitions. The hierarchical structure allows for the sharing of information across groups while still accommodating individual variations in the partition structures.

\section{Posterior Inference}\label{sec:post}
\noindent Posterior inference for the parameters of the LLDPM is carried out using a Markov chain Monte Carlo (MCMC) algorithm and, more specifically, a Gibbs sampling scheme. 
For computational convenience, we marginalize with respect to the local level parameters $\bm{\beta}_t$, thus introducing, for any $t=1,\ldots,T$, the conditional distribution of the $n$-dimensional vector of observations $\bm{Y}_t$, 
given the partition $\pi_t$, namely
\begin{equation*}
p(\bm{Y}_t\mid \pi_t) = \prod_{j=1}^{\mid \pi_t\mid} \int \prod_{i \in C_{j,t}} p(y_{i,t}\mid \beta_{j,t}^\ast) P_0(\d\beta_{j,t}^\ast).
\end{equation*}
The algorithm comprises three main steps, including a joint update of the random partition $\pi_t$ and the changepoint $\gamma_t$ from their full conditional distribution. The updates also include the dependence parameters $\eta_t$, for which we specify a hyperprior distribution as $\eta_t \simiid\text{Beta}(a,b)$. The steps of the implemented algorithm are described in Section \ref{appendix:MCMC} of the Supplementary Material, along with additional details on the posterior distributions and our implementation.

Changepoint detection inherently involves making multiple comparisons,  since the decisions are temporally dependent. To address this multi-comparison problem, we use a compound decision-theoretic approach to detect the presence of a changepoint, which is based on a loss function that takes simultaneously into account the sequence of decisions, and it is defined as a linear combination of measures of the false positive and true positive (or, alternatively, false negative) decisions \citep{sun2007}. In a Bayesian context, \citet{muller2004optimal} and \citet{muller2006fdr} have demonstrated that, under the assumption of both independent hypotheses and independent (marginal) loss functions, the optimal approach for minimizing the resulting posterior expected loss involves thresholding the posterior probabilities of the changepoint ($\text{PPC}_t$), $\Pr(\gamma_t = 1 \mid \bm{Y}_{1:T})$, for \(t=2, \ldots, T\), as estimated from the MCMC output. However, such a procedure does not inherently control for a given false discovery rate (FDR) unless such control is explicitly accounted for. That is, we need to determine the optimal threshold in order to control the FDR at a specific desired level, say $\zeta$. More in detail, we consider the Bayesian FDR \citep{newton2004detecting},
\begin{equation}
    \text{BFDR}_{\text{m}}(h) = \frac{\sum_{t=2}^T (1-\text{PPC}_t) \mathbbm{1}_{\{\text{PPC}_t>h\}}}{\max\left(\sum_{t=2}^T \mathbbm{1}_{\{\text{PPC}_t>h\}},1\right)},
    \label{eq:FDR}
\end{equation}
where $h$ is the chosen threshold and $\mathbbm{1}_{\{\text{PPC}_t>h\}}$ is an  indicator function such that $\mathbbm{1}_{\{\text{PPC}_t>h\}} = 1$ if $\text{PPC}_t>h$, and 0 otherwise. The optimal threshold $h^{*}$ corresponds
to the minimum value of $h$ that ensures that the $\mathrm{BFDR}_{\mathrm{m}}$ is less than $\zeta$.  In formulas, $h^*=\min \{h: \operatorname{BFDR}_{\text{m}}(h) \leq \zeta\}$. 
The previous testing procedure can be classified as a \emph{marginal} approach (hence, the subscript in $\text{BFDR}_{\text{m}}$) since it fails to consider existing dependence either among hypotheses or in the decisions themselves.
 \citet{sun2015} extended this framework to the spatial setting, explicitly accounting for dependence among hypotheses, as induced by a spatial model. More recently, \citet{chandra2019non} introduced \emph{non-marginal} loss functions and non-marginal decision rules that directly account for dependence in the decision-making process. Their procedure incorporates additional information about dependence among tests into the definition of error and non-error terms for subgroups of hypotheses. Specifically, the approach penalizes each hypothesis based on incorrect decisions related to other dependent tests, thus defining a compound loss where decisions about dependent tests are interrelated. We adapt their framework to our setting, and the resulting changepoint detection procedure is presented in Section \ref{sec:changepoint} of the Supplementary Material.
 
\section{Simulation studies}\label{sec:sim}

We present two simulation studies to illustrate the performance of our LLDPM under different data-generating mechanisms. We consider two versions of the LLDPM, defined by the choice of base partition distribution. In the first version, denoted LLDPM$_0$, we set \( p^*(\cdot) = p_{\text{CRP}}(\cdot) \), corresponding to a CRP. In the second version, denoted LLDPM$_{0.25}$, we use a 2-CRP and set \( p^*(\cdot) = p_{\sigma}(\cdot) \) with a discount parameter \( \sigma = 0.25 \). Such choice of \( \sigma \) is arbitrary and simply intended to demonstrate the applicability of our method when the base partition distribution differs from the CRP. While other choices are possible, additional investigations, not reported here, suggest that the method is rather robust to such variations.  
We then investigate the models ability to accurately detect changepoints in both independent and autocorrelated data scenarios, while also recovering the time-specific latent cluster structure. For comparison, we evaluate our models against six alternatives: (1) the Dependent Random Partition Model (DRPM) introduced by \citet{page2022dependent}, which, to our knowledge, is the only model-based approach that introduces time dependence directly through partitions; (2) the Linear Dependent Dirichlet Process (LDDP)  described by \citet{de2004anova}, which incorporates time explicitly within the atoms of the dependent process; (3) the Weighted Dependent Dirichlet Process (WDDP) by \citet{quintana2022dependent}, which integrates time into the weights of the Dirichlet process; (4) the Griffiths--Milne Dependent Dirichlet Process (GMDDP) introduced by \citet{lijoi2014bayesian} and implemented in the \texttt{R} package \texttt{BNPmix} \citep{corradin2021bnpmix}, which defines multivariate vectors of dependent and identically distributed Dirichlet processes; (5) the Product Partition Model (PPM) implemented in the \texttt{R} package \texttt{ppmSuite} \citep{page2022ppmsuite}, a Bayesian model where the set of changepoint indicators between time series are correlated; (6) the Autoregressive Dirichlet Process (AR1-DP) introduced by \citet{de2023bayesian}, which adopts an autoregressive prior to allow for temporal dynamic clustering.

It is important to emphasize that among the six alternatives considered, only the DRPM is explicitly designed to model the evolution of random partitions, as is the case for our proposed model. The remaining models were not devised for this purpose. Specifically, the PPM  is designed for multivariate time series and detects changepoints as shifts in the mean of the process rather than changes in partitions. 
The LDDP, WDDP, and GMDDP models are adapted for changepoint detection by organizing the data as a single vector and incorporating time as a covariate, following the approach of \citet{page2022dependent}. The AR1-DP considers a sequence of autoregressive random probability measures over time, but does not explicitly incorporate changepoints detection. To identify changepoints when using models that are not specifically designed to detect changes in the partition evolution over time, we rely on a similarity-based criterion that compares cluster configurations between consecutive time points. Given that recording a changepoint any time the two partitions are not identical would lead to an overidentification of changepoints, our approach allows for a more flexible assessment of partition similarity. This criterion is applied to the competitor models DRPM, LDDP, WDDP, GMDDP, and AR1-DP. Specifically and only for these methods, a changepoint is identified when the ARI between consecutive partitions falls below 85\%, in combination with the same FDR correction adopted for our LLDPM. This adjustment leads to improved performance for these methods (see Supplementary Material, Section \ref{sm:metrics}, for details). For the PPM model, we adopt a different strategy and rely on the estimated individual changepoints, as the model does not account for time-specific partitions. \\
We evaluate the performance of the above methods on two key tasks: accurately estimating the true partition of subjects at each time point and correctly identifying changepoints in partition evolution over time. For the first task, we use the ARI to compare the true partition with the one estimated by minimizing the lower bound of the posterior expected Variation of Information \citep[VI;][]{wade2018bayesian}. For the second task, we use six standard metrics, namely specificity, accuracy, recall, precision, F1-score, and AUC, as defined in Section \ref{sm:metrics} of the Supplementary Material. For the PPM model, these metrics are first computed at the individual changepoint level and then averaged across the entire sample. 
Finally, due to its autoregressive structure, the AR1-DP is applied exclusively in the autoregressive data 
scenario of Section \ref{sec:ar1_data}.

\subsection{Simulations with independent data}
\label{sec:indep}

\begin{table}[t!]
\begin{adjustbox}{max width=\textwidth}
\begin{tabular}{clccccccc}
\hline
$n$ & Measure & LLDPM$_0$ & LLDPM$_{0.25}$ & DRPM & LDDP & WDDP & GMDDP & PPM \\
\hline
\multirow{6}{*}{20}
& specificity & 0.99 (0.01) & 0.97 (0.03) & 0.99 (0.01) & 0.91 (0.17) & 0.47 (0.04) & $\mathbf{1.00}$ (0.00) & 0.77 (0.01) \\
& accuracy    & 0.99 (0.01) & 0.97 (0.03) & 0.96 (0.01) & 0.92 (0.16) & 0.52 (0.03) & $\mathbf{1.00}$ (0.00) & 0.72 (0.01) \\
& recall      & $\mathbf{1.00}$ (0.00) & $\mathbf{1.00}$ (0.00) & 0.63 (0.14) & $\mathbf{1.00}$ (0.00) & 1.00 (0.02) & $\mathbf{1.00}$ (0.00) & 0.23 (0.01) \\
& precision   & 0.90 (0.09) & 0.76 (0.16) & 0.89 (0.12) & 0.81 (0.32) & 0.14 (0.01) & $\mathbf{1.00}$ (0.00) & 0.08 (0.00) \\
& F1          & 0.95 (0.05) & 0.85 (0.11) & 0.73 (0.12) & 0.85 (0.28) & 0.25 (0.01) & $\mathbf{1.00}$ (0.00) & 0.13 (0.01) \\
& AUC         & 0.99 (0.01) & 0.98 (0.01) & 0.80 (0.08) & 0.85 (0.21) & 0.73 (0.05) & $\mathbf{1.00}$ (0.00) & 0.50 (0.01) \\
\hline
\multirow{6}{*}{50}
& specificity & 0.96 (0.03) & 0.91 (0.08) & 0.99 (0.01) & 0.44 (0.40) & 0.29 (0.10) & $\mathbf{1.00}$ (0.03) & 0.66 (0.00) \\
& accuracy    & 0.97 (0.02) & 0.91 (0.07) & 0.96 (0.36) & 0.49 (0.36) & 0.35 (0.09) & $\mathbf{1.00}$ (0.02) & 0.63 (0.00) \\
& recall      & $\mathbf{1.00}$ (0.00) & $\mathbf{1.00}$ (0.00) & 0.67 (0.15) & $\mathbf{1.00}$ (0.00) & 1.00 (0.02) & $\mathbf{1.00}$ (0.00) & 0.34 (0.00) \\
& precision   & 0.73 (0.15) & 0.56 (0.23) & 0.88 (0.11) & 0.34 (0.38) & 0.11 (0.01) & $\mathbf{0.99}$ (0.09) & 0.08 (0.00) \\
& F1          & 0.84 (0.10) & 0.70 (0.18) & 0.74 (0.10) & 0.41 (0.35) & 0.20 (0.02) & $\mathbf{0.99}$ (0.07) & 0.14 (0.00) \\
& AUC         & 0.98 (0.01) & 0.91 (0.14) & 0.83 (0.07) & 0.68 (0.20) & 0.64 (0.05) & $\mathbf{1.00}$ (0.01) & 0.50 (0.00) \\
\hline
\multirow{6}{*}{100}
& specificity & 0.95 (0.03) & 0.83 (0.10) & $\mathbf{0.99}$ (0.01) & 0.38 (0.36) & 0.23 (0.04) & 0.96 (0.11) & 0.58 (0.00) \\
& accuracy    & 0.95 (0.03) & 0.85 (0.09) & $\mathbf{0.96}$ (0.02) & 0.43 (0.33) & 0.28 (0.04) & 0.28 (0.10) & 0.56 (0.00) \\
& recall      & $\mathbf{1.00}$ (0.00) & $\mathbf{1.00}$ (0.00) & 0.60 (0.15) & $\mathbf{1.00}$ (0.00) & 0.88 (0.03) & $\mathbf{1.00}$ (0.00) & 0.44 (0.00) \\
& precision   & 0.66 (0.17) & 0.40 (0.16) & 0.81 (0.15) & 0.26 (0.32) & 0.09 (0.01) & $\mathbf{0.92}$ (0.24) & 0.08 (0.00) \\
& F1          & 0.78 (0.12) & 0.55 (0.16) & 0.68 (0.13) & 0.34 (0.30) & 0.16 (0.01) & $\mathbf{0.93}$ (0.20) & 0.14 (0.00) \\
& AUC         & 0.97 (0.02) & 0.83 (0.17) & 0.79 (0.08) & 0.67 (0.18) & 0.55 (0.02) & $\mathbf{0.98}$ (0.06) & 0.52 (0.00) \\
\hline
\end{tabular}
\end{adjustbox}
\caption{Section \ref{sec:indep}. Performance measures for changepoint detection with independent data, for the LLDPM under two specifications and five competing models. The values correspond to the average (standard deviation) over 50 replicates.}
\label{tab:summ_stat}
\end{table}

We begin by simulating independent data characterized by similar partition structures over different intervals of time, trying to mimic the data-generating process characterizing the LLDPM. We evaluate performance across 50 replicated datasets and different numbers of observations, specifically \(n = \{20, 50, 100\}\), over \(T = 100\) time points. In each scenario, we consider data partitions characterized by eight changepoints. 
The time axis is divided into nine consecutive blocks of varying lengths, each representing a segment of the 100 total time points. Within each block, the partition of the units remains fixed and corresponds to one of three configurations. Two of these configurations consist of three clusters and differ only in the assignment of units to clusters, while the third configuration includes just two clusters. These configurations alternate across blocks. At each time point, given the time-specific partition defined by the active configuration, observations for each cluster are drawn from a Normal distribution. The variance is fixed at 0.01 and shared across all clusters, while the mean for each cluster is independently sampled from a Normal distribution with arbitrarily chosen parameters. 
An example of the generated data is shown in Figure \ref{fig:indep_data} in the Supplementary Material.

To analyze these data, the specification of the two versions of our LLDPM is completed 
by assuming $\theta=1$ and that the base probability distribution $P_0(\cdot)$ is Normal with mean zero and variance $\varsigma^2$.  We further specify the following hyperprior distributions: $\tau^2 \sim \text{Inv-Gamma}(15,3)$ and $\varsigma^2 \sim \text{Inv-Gamma}(15,3)$. Additionally, we set $\eta_t \simiid \text{Beta}(0.1,0.9)$, which places greater probability on the absence of a changepoint at each time point.  When fitting the DRPM model, we set the parameters as suggested in \citet{page2022dependent}, except for the variances, which we fixed as in our model. The temporal dependence parameter in the DRPM is modeled as the analogous parameters $\eta_t$ in the LLDPM. Given the different parameterizations of the two models, this corresponds to a \(\text{Beta}(0.9, 0.1)\) distribution. Finally, for all the other considered models, we used the same variances as in our model and specified the default hyperprior parameter values.\\
We implemented the Gibbs sampling scheme detailed in Section \ref{sec:Gibbs} of the Supplementary Material, running the algorithm for 10,000 iterations and discarding the first 5,000 as burn-in. We randomly selected a subset of replications and visually inspected the trace plots of the time-specific number of clusters and the partition entropy. These diagnostics revealed no signs of convergence issues. Examples of these trace plots are provided in Section \ref{sm:plots} of the Supplementary Material.  Subsequently, we calculated the optimal partition for each time point based on the estimated posterior similarity matrix. To determine whether a specific time point $t$ should be considered a changepoint, we employed the Bayesian False Discovery Rate method \citep{newton2004detecting, muller2006fdr}. As explained in Section \ref{sec:changepoint} of the Supplementary Material, we implemented a penalized version of FDR with a control level set at 0.01/3.

\begin{figure}[t!]
    \centering
    \includegraphics[clip, trim=0cm 0cm 0cm 0.15cm, width=0.85\textwidth]{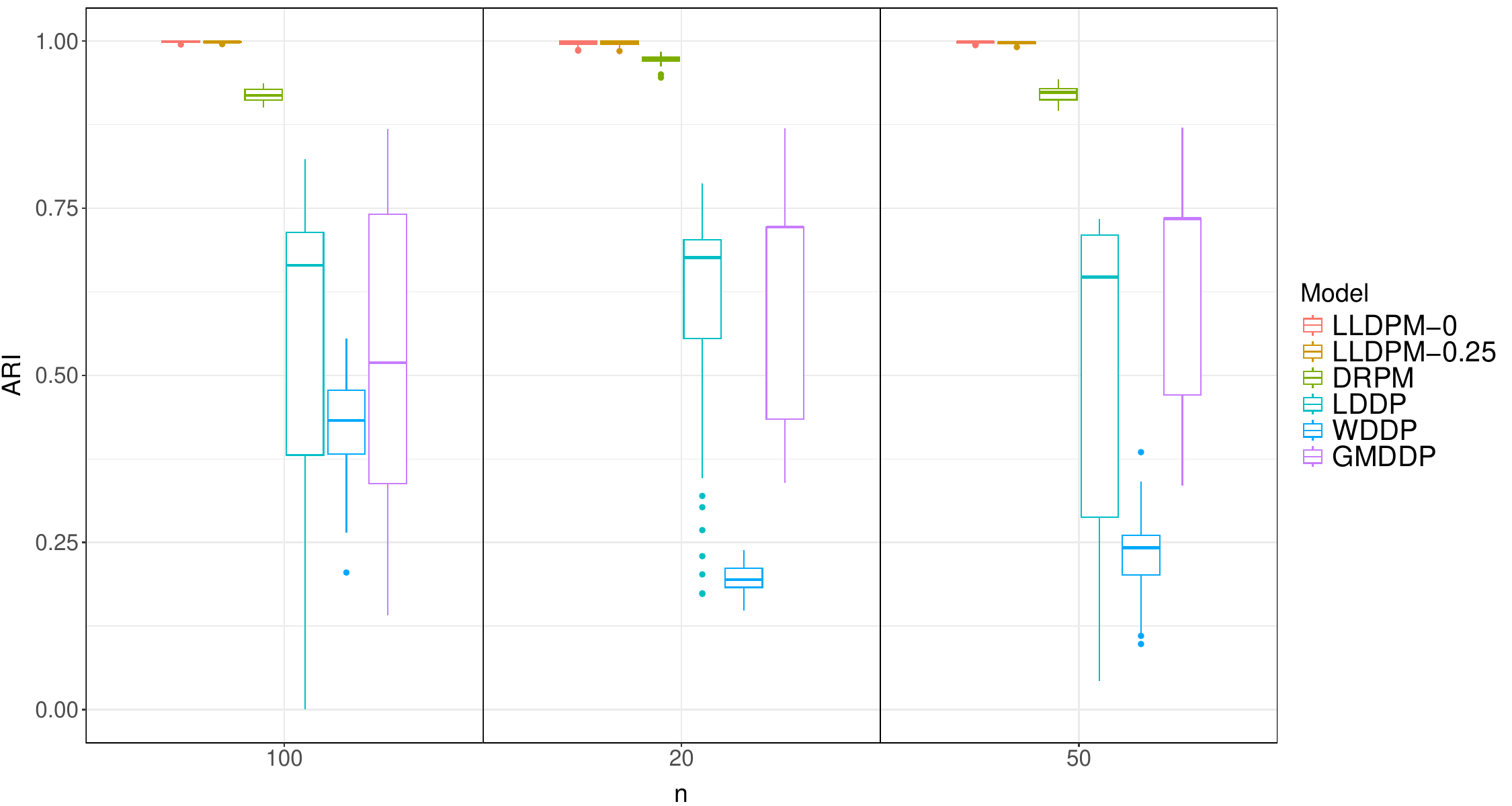}
    \caption{Section \ref{sec:indep}. Boxplots of the average ARI, to evaluate the clustering performance when analyzing independent data,  with $n=\{20,50,100\}$, for two versions of our LLDPM model and four alternatives. The PPM is not included in the comparison as the model does not account for time-specific partitions of the units. Results are based on 50 replicated datasets and averaged over 100 time points. For each dataset and each method, at each time point, the true partition is compared with the partition estimated by minimizing the lower bound of the posterior expected VI.}
    \label{fig:ari_ind_data}
\end{figure}

Table \ref{tab:summ_stat} summarizes the performance of the seven models considered, including our LLDPM$_{0}$ and LLDPM$_{0.25}$, and models 1–5 introduced at the beginning of Section \ref{sec:sim}, across varying data dimensions $n$. Notably, both versions of our model exhibit strong performance, consistently outperforming LDDP, WDDP, and PPM. All changepoint detection metrics highlight the robustness of LLDPM, displaying strong and stable results regardless of the underlying base process specification, whether based on a CRP or a 2-CRP with $\sigma=0.25$. Further remarks are needed for the remaining two competitors, GMDDP and DRPM. While GMDDP performs well in detecting changepoints, as reflected in the evaluation metrics, it struggles to accurately recover the true time-specific latent cluster structures, as illustrated in Figure \ref{fig:ari_ind_data}. Consequently, GMDDP may not be well-suited for applications where both changepoint detection and cluster analysis are of interest. Also DRPM shows strong results in changepoint detection, achieving specificity levels comparable to those of LLDPM across all settings. However, it suffers from lower recall, indicating a tendency to miss many true changepoints. In contrast, LLDPM excels in both tasks, consistently delivering high changepoint detection accuracy and near-perfect ARI scores for latent cluster recovery, see Figure \ref{fig:ari_ind_data}. 
Finally, we observe that as the sample size increases, performance tends to decline across all models, due to the greater likelihood of subjects switching clusters.\\
To evaluate the performance of our model for sample sizes larger than those considered in Table \ref{tab:summ_stat}, we used both versions of the LLDPM to analyze 50 replicated datasets with $n=500$ observations over $T=100$ time points. Due to memory limitations and prohibitively long runtimes, the competing models could not be fitted. 
In contrast, for our models, we adopted a blockwise Gibbs sampling strategy to manage the computational load, splitting the sampling into smaller blocks of iterations to prevent memory and performance issues. Both LLDPM$_{0}$ and LLDPM$_{0.25}$ successfully identified the changepoints, achieving perfect recall. However, their performance in correctly identifying periods with no-changepoint declined, leading to an increase in  false-positives. Specifically, LLDPM$_{0}$ achieved over 70\% specificity and accuracy,  with an AUC close to 80\%. However, its F1 score dropped to around 40\% due to lower precision.  LLDPM$_{0.25}$ showed a more 
polarized behavior with specificity and accuracy around 40\%, an AUC of 70\%, and an F1 score of about 25\%. Despite these differences in changepoint detection metrics, both models achieved an average ARI of approximately $99.95\%$, indicating a near-perfect recovery of the time-specific latent cluster structures.

\subsection{Simulations with autoregressive data}
\label{sec:ar1_data}

\begin{table}[t!]
\begin{adjustbox}{max width=\textwidth}
\begin{tabular}{clcccccccc}
\hline
$\lambda$ & Measure & LLDPM$_0$ & LLDPM$_{0.25}$ & DRPM & LDDP & WDDP & GMDDP & PPM & AR1-DP \\
\hline
\multirow{6}{*}{0.25}
& specificity & 0.91 (0.08) & 0.95 (0.06) & 0.98 (0.04) & $\mathbf{1.00}$ (0.00) & 0.82 (0.05) & 0.87 (0.10) & 0.83 (0.02) & 0.64 (0.15) \\
& accuracy    & 0.91 (0.06) & 0.94 (0.04) & 0.52 (0.04) & $\mathbf{1.00}$ (0.02) & 0.52 (0.03) & 0.65 (0.07) & 0.49 (0.00) & 0.83 (0.07) \\
& recall      & 0.92 (0.07) & 0.94 (0.06) & 0.13 (0.07) & 1.00 (0.04) & 0.25 (0.05) & 0.45 (0.07) & 0.21 (0.02) & $\mathbf{1.00}$ (0.00) \\
& precision   & 0.93 (0.07) & 0.96 (0.05) & 0.91 (0.16) & $\mathbf{1.00}$ (0.00) & 0.62 (0.09) & 0.81 (0.12) & 0.61 (0.01) & 0.77 (0.08) \\
& F1          & 0.92 (0.06) & 0.95 (0.04) & 0.21 (0.11) & $\mathbf{1.00}$ (0.02) & 0.35 (0.06) & 0.58 (0.08) & 0.31 (0.02) & 0.87 (0.05) \\
& AUC         & 0.91 (0.06) & 0.94 (0.04) & 0.54 (0.04) & $\mathbf{1.00}$ (0.02) & 0.53 (0.03) & 0.66 (0.07) & 0.52 (0.00) & 0.82 (0.08) \\
\hline
\multirow{6}{*}{0.5}
& specificity & 0.88 (0.10) & 0.94 (0.07) & 0.96 (0.18) & $\mathbf{0.99}$ (0.02) & 0.80 (0.06) & 0.67 (0.16) & 0.82 (0.02) & 0.55 (0.14) \\
& accuracy    & 0.89 (0.07) & 0.93 (0.05) & 0.51 (0.04) & $\mathbf{1.00}$ (0.01) & 0.51 (0.04) & 0.58 (0.08) & 0.49 (0.00) & 0.79 (0.07) \\
& recall      & 0.90 (0.08) & 0.93 (0.07) & 0.13 (0.13) & $\mathbf{1.00}$ (0.00) & 0.25 (0.06) & 0.51 (0.07) & 0.22 (0.03) & $\mathbf{1.00}$ (0.00) \\
& precision   & 0.90 (0.07) & 0.95 (0.06) & 0.84 (0.21) & $\mathbf{0.99}$ (0.02) & 0.59 (0.08) & 0.65 (0.12) & 0.60 (0.01) & 0.72 (0.07) \\
& F1          & 0.90 (0.06) & 0.94 (0.05) & 0.21 (0.11) & $\mathbf{1.00}$ (0.01) & 0.35 (0.07) & 0.57 (0.07) & 0.32 (0.02) & 0.84 (0.04) \\
& AUC         & 0.89 (0.07) & 0.93 (0.05) & 0.54 (0.04) & $\mathbf{1.00}$ (0.01) & 0.52 (0.04) & 0.60 (0.07) & 0.52 (0.00) & 0.78 (0.07) \\
\hline
\multirow{6}{*}{0.75}
& specificity & 0.86 (0.12) & 0.92 (0.08) & 0.94 (0.07) & $\mathbf{0.96}$ (0.08) & 0.77 (0.07) & 0.45 (0.25) & 0.83 (0.02) & 0.34 (0.14) \\
& accuracy    & 0.87 (0.06) & 0.92 (0.06) & 0.52 (0.04) & $\mathbf{0.98}$ (0.04) & 0.49 (0.04) & 0.66 (0.16) & 0.49 (0.00) & 0.69 (0.07) \\
& recall      & 0.88 (0.08) & 0.91 (0.08) & 0.16 (0.08) & $\mathbf{1.00}$ (0.00) & 0.24 (0.06) & 0.84 (0.14) & 0.22 (0.02) & 1.00 (0.01) \\
& precision   & 0.89 (0.09) & 0.93 (0.07) & 0.79 (0.17) & $\mathbf{0.97}$ (0.06) & 0.55 (0.09) & 0.65 (0.14) & 0.61 (0.01) & 0.64 (0.05) \\
& F1          & 0.88 (0.06) & 0.92 (0.06) & 0.25 (0.10) & $\mathbf{0.98}$ (0.03) & 0.33 (0.07) & 0.73 (0.12) & 0.32 (0.02) & 0.78 (0.04) \\
& AUC         & 0.87 (0.06) & 0.92 (0.06) & 0.53 (0.04) & $\mathbf{0.98}$ (0.04) & 0.51 (0.04) & 0.65 (0.16) & 0.52 (0.00) & 0.67 (0.07) \\
\hline
\multirow{6}{*}{0.9}
& specificity & 0.85 (0.10) & 0.90 (0.10) & $\mathbf{0.95}$ (0.10) & 0.49 (0.27) & 0.77 (0.09) & 0.37 (0.31) & 0.81 (0.03) & 0.22 (0.10) \\
& accuracy    & 0.87 (0.06) & $\mathbf{0.90}$ (0.06) & 0.51 (0.04) & 0.75 (0.13) & 0.49 (0.04) & 0.65 (0.16) & 0.49 (0.01) & 0.62 (0.05) \\
& recall      & 0.88 (0.06) & 0.90 (0.07) & 0.13 (0.08) & $\mathbf{0.98}$ (0.04) & 0.24 (0.07) & 0.91 (0.19) & 0.23 (0.03) & 0.97 (0.04) \\
& precision   & 0.89 (0.08) & $\mathbf{0.91}$ (0.07) & 0.79 (0.17) & 0.70 (0.12) & 0.55 (0.12) & 0.63 (0.18) & 0.60 (0.01) & 0.59 (0.03) \\
& F1          & 0.88 (0.07) & $\mathbf{0.90}$ (0.05) & 0.22 (0.10) & 0.81 (0.08) & 0.33 (0.08) & 0.75 (0.12) & 0.33 (0.03) & 0.73 (0.03) \\
& AUC         & 0.86 (0.06) & $\mathbf{0.90}$ (0.06) & 0.52 (0.05) & 0.73 (0.14) & 0.50 (0.04) & 0.65 (0.15) & 0.52 (0.00) & 0.60 (0.05) \\
\hline
\end{tabular}
\end{adjustbox}
\caption{Section \ref{sec:ar1_data}.  
Performance measures for changepoint detection with AR(1) data for two versions of our LLDPM model and six alternatives. Values correspond to the average (standard error) over 50 replicates.}
\label{tab:lambdas}
\end{table}

In a second simulation study, we aim to assess the performance of the two versions of our LLDPM in scenarios where the data are generated from a process characterized by an autoregressive structure of order one, or AR(1). Specifically, we consider datasets with $T=30$ time points and $n=20$ units, generated from the model
\begin{equation}\label{eq:AR1data} 
Y_{i,t} = \lambda Y_{i,t-1} + \tilde{\beta}_{i,t} + \tilde{\varepsilon}_{i,t},
\end{equation}
where the error terms \(\tilde{\varepsilon}_{i,t}\) are assumed to be independent and identically distributed from a Normal with mean zero and variance one. At each time point \(t=1,\ldots,T\), units are partitioned based on the distinct values taken by \(\tilde{\beta}_{i,t}\). The number of distinct values, and thus the number of clusters, varies between one and two over time. Specifically, there are two clusters when \(t\) is divisible by five or nine, and one cluster otherwise. When \(t\) is divisible by five, the two clusters are of equal size; when $t$ is divisible by nine, one cluster is more than twice the size of the other. This setup gives rise to three distinct configurations that alternate across time points, with the allocation of units to clusters updated whenever the configuration changes. The observations are thus simulated from model \eqref{eq:AR1data}, with the values of $\tilde{\beta}_{i,t}$ at each time point assigned to arbitrarily fixed, cluster-specific values. 
An example of such data is shown in Figure \ref{fig:ar1_res_cgp_lambda0.9} in the Supplementary Material. When fitting the two versions of our model, unlike in Section \ref{sec:indep}, we set \( \theta \) based on the expected prior number of clusters, \( \mathbb{E}[|\pi_t|] \), for each time point \( t = 1, \ldots, T \). For the LLDPM with specification \(p^*(\cdot) = p_{\text{CRP}}(\cdot)\), this is given by \(\mathbb{E}[|\pi_t|] = \sum_{i=1}^n \theta/\theta + i - 1\), while for the LLDPM with \(p^*(\cdot) = p_{\sigma}(\cdot)\) and $\sigma\in(0,1)$, by \(\mathbb{E}[|\pi_t|] = \left[(\theta + \sigma)_n / (\theta + 1)_{n-1} - \theta\right]/\sigma\) \citep{pitman2006combinatorial}. 
By setting \(\mathbb{E}[|\pi_t|] = 2\) for all \(t = 1, \ldots, T\), we obtain \(\theta = 0.32\) for LLDPM$_0$ and \(\theta = -0.07\) for LLDPM$_{0.25}$. Moreover, we let the autoregressive parameter $\lambda $ vary in $\{0.25, 0.5, 0.75, 0.9\}$ to see if increasing the dependence across times affect the performances of the model. For the other parameters, we use the same specifications as in the simulation study of Section \ref{sec:indep}. As before, we run the MCMC for 10,000 iterations, discarding the first half as burn-in. 

\begin{figure}[t!]
    \centering
    \includegraphics[clip,trim=0cm 0cm 0cm 0.15cm,width=0.85\textwidth]{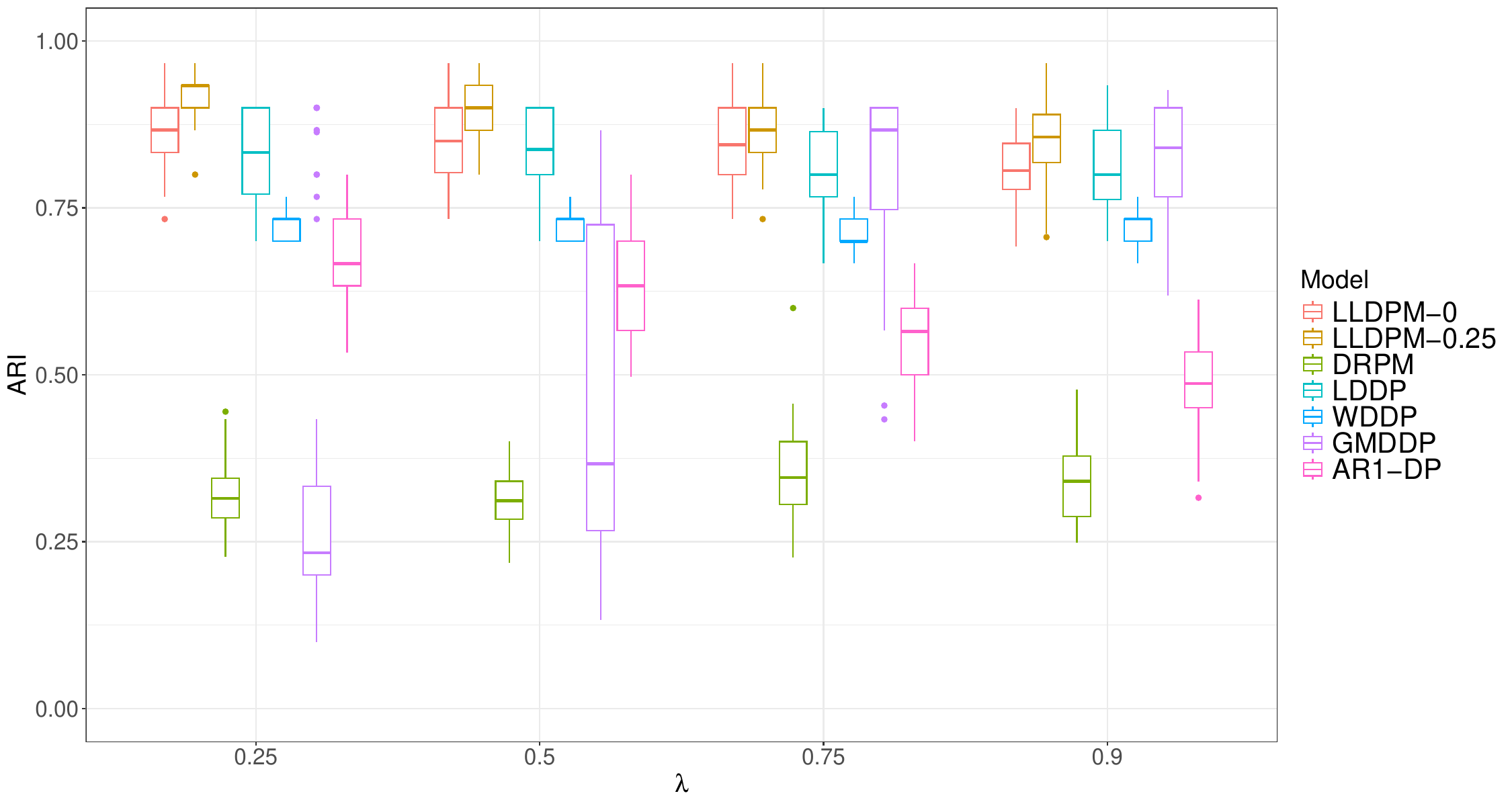}
    \caption{Section \ref{sec:ar1_data}. Boxplots of the average ARI, %are presented 
    to evaluate clustering performance when analyzing AR(1) data, comparing two versions of our model with five alternatives. The PPM is excluded from the comparison as it does not account for time-specific partitions of the units. The analysis considers different values of the autoregressive coefficient \(\lambda = \{0.25, 0.5, 0.75, 0.9\}\). Results are based on 50 replicated datasets and averaged over 30 time points. For each dataset and each method, at each time point, the true partition is compared with the partition estimated by minimizing the lower bound of the posterior expected VI. }   \label{fig:ari_data_ar1_nocrp-label}
\end{figure}

The results of our analysis, based on 50 replicates, are presented in Table \ref{tab:lambdas}. Both versions of the LLDPM consistently outperform most other models in detecting changepoints across all considered metrics. In this second study, we observe that GMDDP exhibits reduced effectiveness compared to the previous simulation. Among the alternatives, LDDP stands out for its strong ability to identify changepoints, although its performance declines in terms of specificity, precision, and accuracy as \(\lambda\) increases. In contrast, both versions of the proposed LLDPM demonstrate robustness, maintaining high performance across all metrics and scenarios, regardless of the degree of temporal dependence in the simulated data. This consistency highlights the reliability of the LLDPM across a wide range of settings. Furthermore, the LDDP falls short in accurately recovering the time-specific latent cluster structure. As shown in Figure \ref{fig:ari_data_ar1_nocrp-label}, both versions of the LLDPM consistently achieve ARI indices above 0.7 across datasets generated with varying autoregressive coefficients, clearly outperforming the other methods. Overall, when considering both changepoint detection and cluster structure recovery, our model provides reliable and robust inference across a wide range of scenarios.

\section{Applications}\label{sec:application}

 We apply our LLDPM model, using the specification \( p^*(\cdot) = p_{\text{CRP}}(\cdot) \), to two real-world datasets: first, the Gesture Phase Segmentation data \citep{madeo2013gesture}, and second, data from the Collaborative Perinatal Project \citep[CPP,][]{longnecker2001association}. Gesture Phase Segmentation focuses on identifying and classifying different phases of hand movements, which is an important task in applications such as sign language recognition, human-computer interaction, and motion analysis. In contrast, CPP is a large-scale epidemiological study that investigates prenatal and perinatal factors affecting pregnancy outcomes. These two applications underscore the versatility of our model across different domains: the first, in Section \ref{sec:fine}, demonstrates LLDPM's effectiveness in changepoint detection for time-series data, while the second, presented in Section \ref{sec:cpp}, illustrates its applicability in a multiview analysis setting.

\subsection{Gesture Phase Segmentation data}\label{sec:gesturedata}

We present an analysis of video-recorded data for human gesture segmentation. The goal is to segment videos into distinct phases exhibiting different motion patterns, e.g., to identify time lapses of the video that need to be removed from a clip \citep{parvathy2021development}. More specifically, we employ the Gesture Phase Segmentation dataset, publicly accessible for download at the following URL: \url{https://archive.ics.uci.edu/ml/datasets/gesture+phase+segmentation}. 
While \citet{madeo2013gesture} analyzed this dataset using Support Vector Machines (SVM) to segment, and classify gesture data streams, they did not consider partitioning gesture phases into their components and investigating the temporal dependence between partitions.  
Our goal, in contrast, is to detect the latent cluster structure at each frame while capturing changepoints, using our proposed LLDPM random partition model with time dependence.

The dataset contains sensor data recordings of users recounting comic book stories facing an Xbox Microsoft Kinect\textsuperscript{TM} sensor. The dataset provides scalar velocity and acceleration values over four sensors, placed on the left hand, right hand, left wrist, and right wrist, leading to $n=8$ sensor measurements at regular time intervals (frames). These values were obtained by normalizing hand and wrist positions relative to the head and spine position using a fixed displacement offset of 3 to measure velocity. This offset is a constant adjustment applied to account for the spatial relationship between the sensors and the torso, which helps standardize the measurements and accurately calculate the velocity.  Our analysis is based on a processed version of the data, which we prepare following the approach outlined by \citet{hadj2023bayesian}. 
The preprocessing consists of four main stages. First, we apply a two-point moving average filter to smooth the time series. Second, to reduce the data volume while preserving essential patterns, we downsample the series by selecting every fifth data point, following the approach suggested by \citet{romanuke2021time}. This strategy allows us to handle longer time lags while minimizing the number of parameters and reducing computational complexity. Third, we transform the data using a square root transformation, which \citet{hadj2023bayesian} found to be effective in fitting a Gaussian distribution to the data. Finally, we standardize the time series to ensure uniformity and comparability across the dataset. 
The processed data is illustrated in Figure \ref{fig:gesture_data}. For additional context, see Figure \ref{fig:gesture_both} in the Supplementary Material, which plots the data along with details on the video phases previously identified by \citet{madeo2013gesture}: D (rest position, from the Portuguese ``descanso''), P (preparation), S (stroke), H (hold), and R (retraction).

\begin{figure}[ht]
    \centering\includegraphics[clip,trim=0cm 0cm 0cm 0.15cm,width=0.85\textwidth]{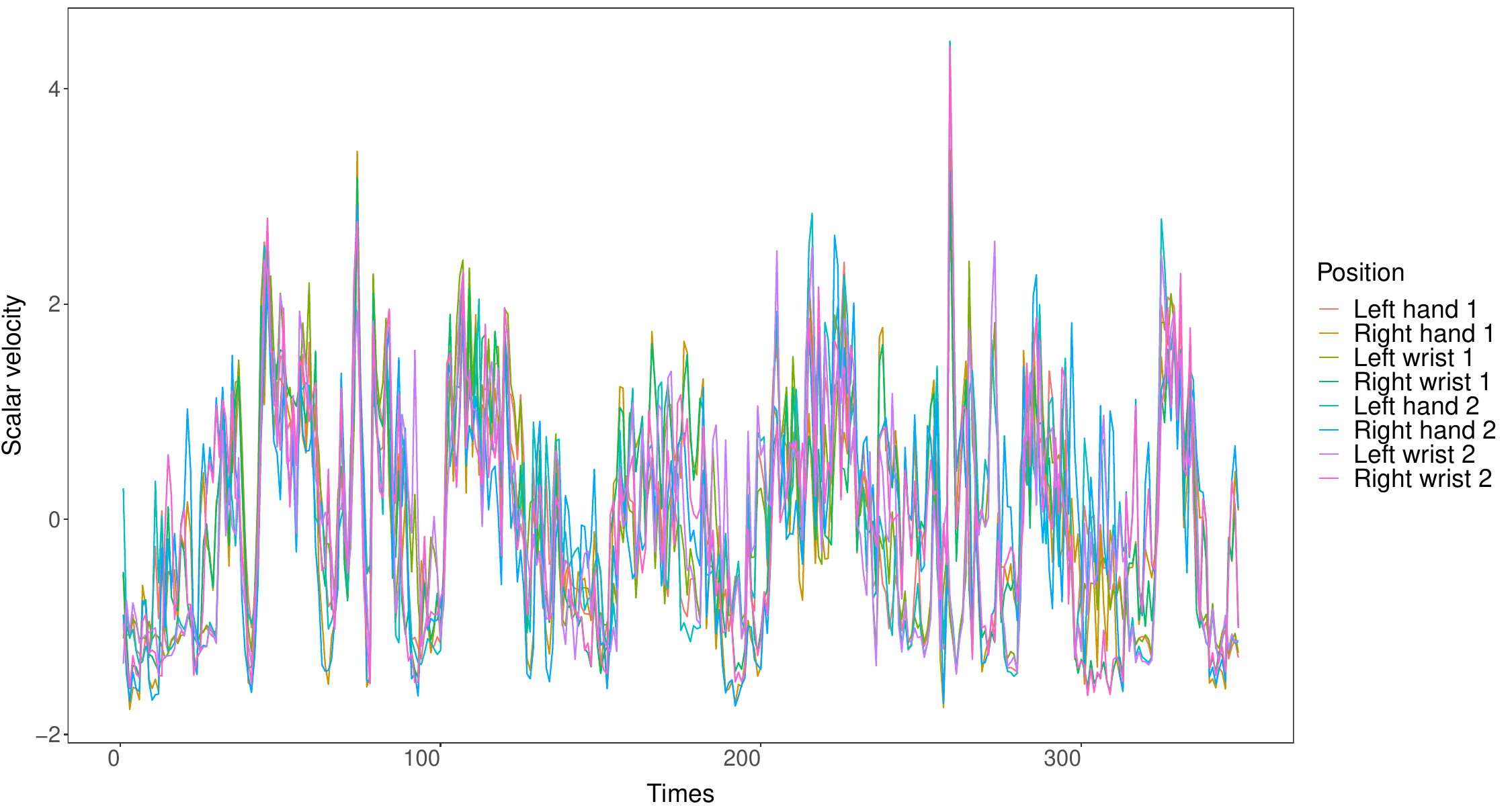}
    \caption{Section \ref{sec:gesturedata}, Human Gesture data. Scalar velocity of the left and right hand and the wrists after preprocessing ($T=349$).}
    \label{fig:gesture_data}
\end{figure}

Here, we illustrate our model's performance in analyzing this data. Gesture phase segmentation poses several challenges. Firstly, it is inherently subjective, with no definitive starting point for each phase, leading to varying segmentations by different experts for the same video. Additionally, similar patterns may occur across different activities; for example, stationary hand gestures may be seen in both the rest and hold phases identified by \citet{madeo2013gesture}. Furthermore, the data may include nuisance movements, such as touching glasses while speaking, which can cause fluctuations in sensor-recorded scalar velocity \citep{madeo2013gesture}.
To implement the LLDPM, we consider an a priori expected number of clusters equal to two at each time point, corresponding to the specification $\theta=0.49$ for the concentration parameter of the CRP. 
The prior probability of a changepoint is assumed $\eta_t \simiid \text{Beta}(0.1,0.9)$, suggesting that the model assumes a relatively low prior probability of a changepoint at each time point. For the distribution $P_0(\cdot)$ of the coefficients $\beta_{i,t}$, we assume a Normal distribution with a mean zero and variance $\varsigma^2 = %0.5^2
1$. Finally, following \citet{corradin2021bnpmix}, we set the variance of the kernel, $\tau^2$, to follow an Inverse-Gamma distribution with a scale parameter equal to the sample variance of the standardized data, i.e., $\tau^2 \sim \text{Inv-Gamma}(2, 1)$. We run our Gibbs sampling algorithm for 10,000 iterations, with the first half discarded as burn-in.  

\begin{figure}[t!]
    \centering
    \includegraphics[clip,trim=0cm 0cm 0cm 0.15cm,width=0.85\textwidth]{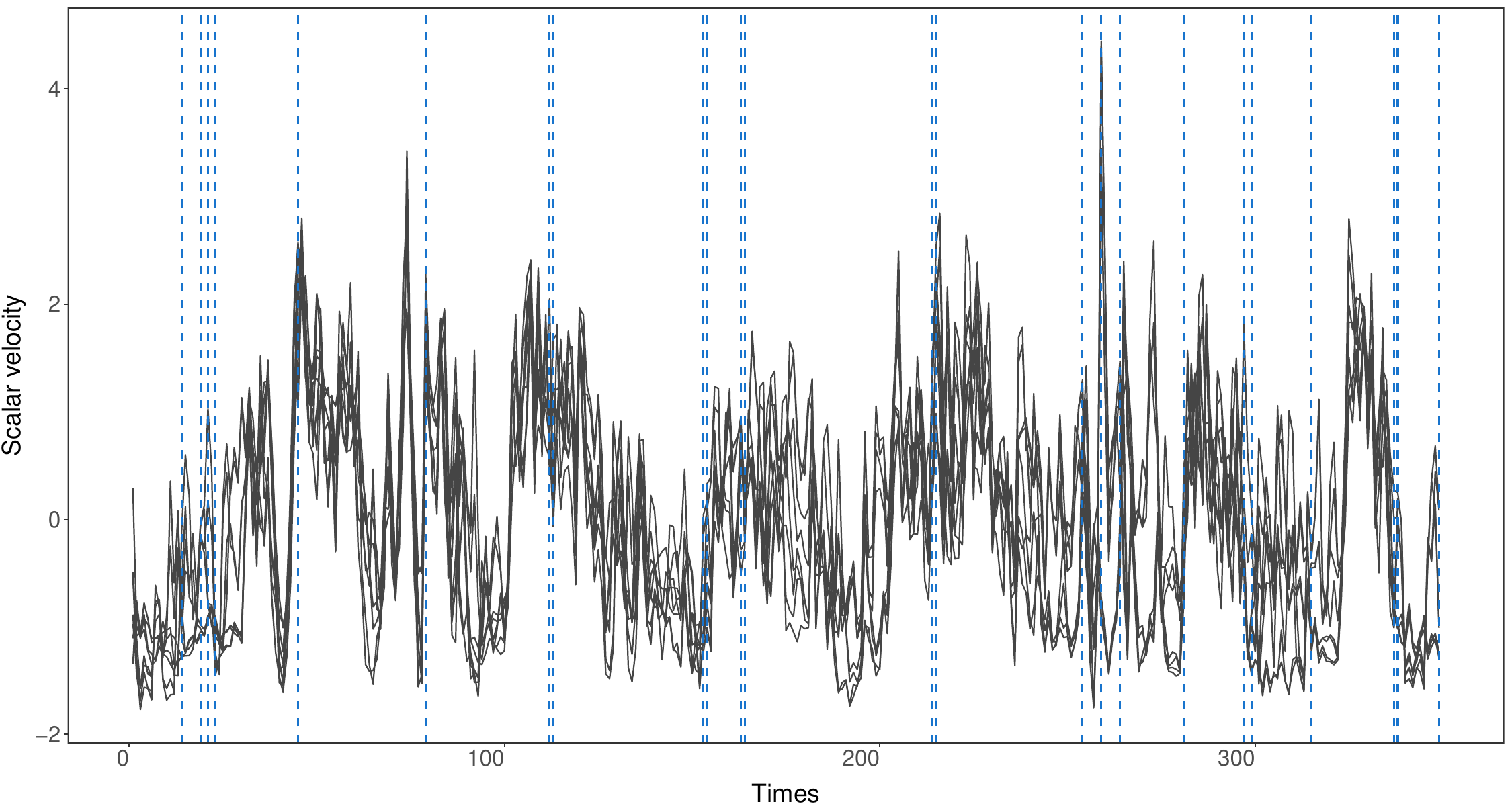}
    \caption{Section \ref{sec:gesturedata}, Human Gesture data.  Changepoints (dashed line) identified with the LLDPM.}
    \label{fig:LLDPM_K2_2phases}
\end{figure}

\begin{figure}[t!]
    \centering
    \includegraphics[clip,trim=0cm 0cm 0cm 0.15cm,width=0.85\textwidth]{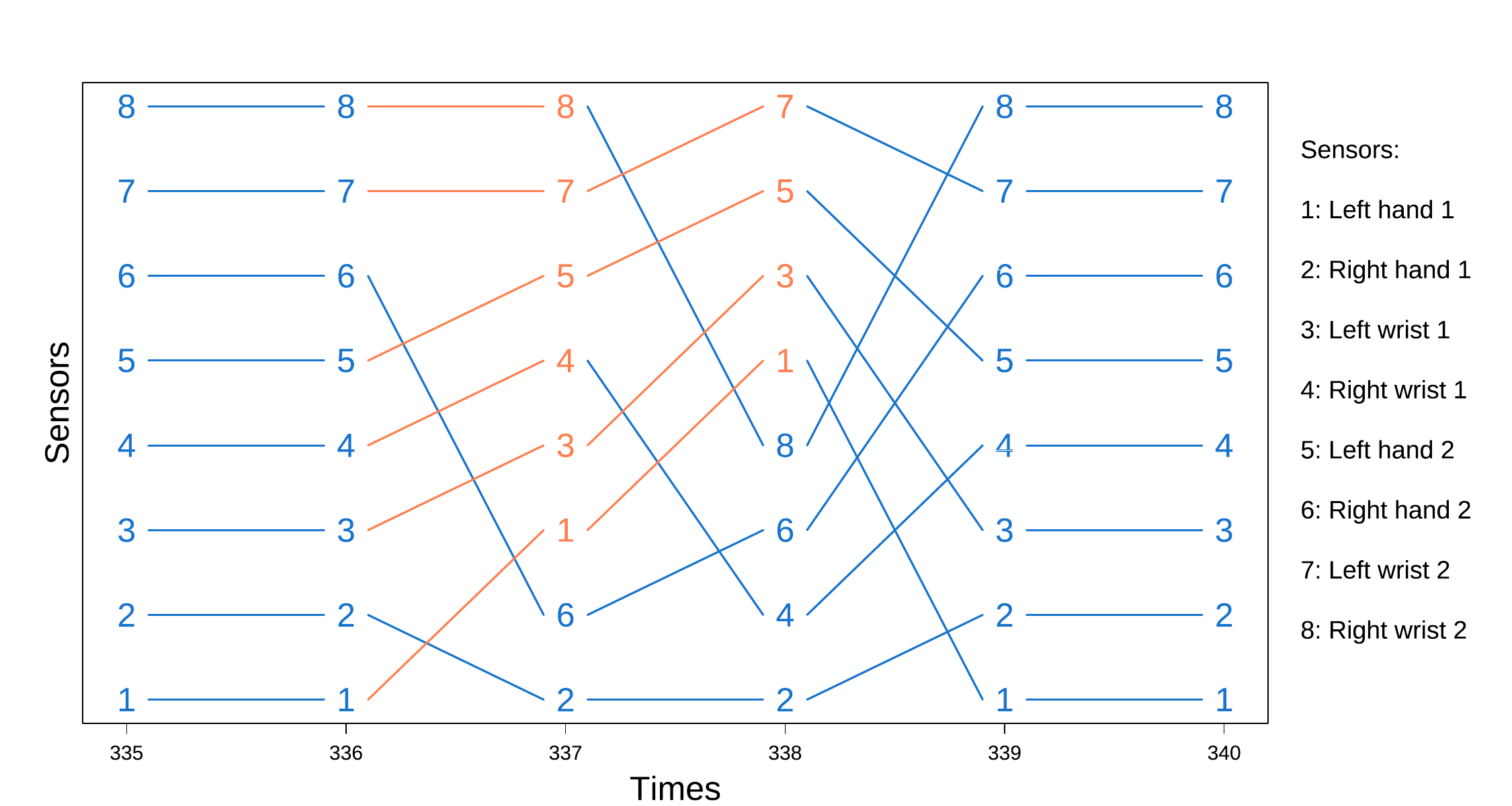}
    \caption{Section \ref{sec:gesturedata}, Human Gesture Data. Time-specific clusters obtained by minimizing the lower bound of the posterior expected VI, %for the time window 
    from time 335 to time 340. Numbers refer to the measurements of scalar velocity (1-4) and acceleration (5-8) in sensor units placed on the hands and wrists of a subject, whereas colors indicate clusters.}
    \label{fig:only_cgp_gesture}
\end{figure}

The estimated changepoints are displayed in Figure \ref{fig:LLDPM_K2_2phases} and appear spread 
throughout the time series, possibly indicating that the person wearing the sensors is gesticulating while recalling the comic story. When compared with the segmentation in \citet{madeo2013gesture}, the changepoints identified by the LLDPM primarily occur during the activity phase, which correspond to the preparation, retraction, and stroke phases in the video. Additionally, most of these changepoints are observed during transitions between different phases, with a notable increase in frequency during the reading phases. These changepoints likely reflect movements related to the subject's gesticulation and body language while narrating, as well as shifts in hand and wrist positions during the assigned activity. 
Figure \ref{fig:only_cgp_gesture} illustrates the clustering of the eight sensors within a specific time window, chosen for representation purposes. The estimated clustering is obtained by minimizing the lower bound of the posterior expected VI. 
The numbers on the plot correspond to the sensor measurements. Notably, a distinct pattern allocation emerges, where all the sensors are grouped into a single primary cluster. Additionally, a secondary pattern takes shape, forming two clusters: one containing all the sensors from the left arm (odd numbers) and the other with sensors from the right arm (even numbers) grouped together.

For comparison, we analyzed the same data with four of the alternative models considered in this work. The identified changepoints are displayed in Figures \ref{fig:drpm_K2_2phases}, \ref{fig:lddp_K2_2phases}, \ref{fig:wddp_K2_2phases} and \ref{fig:gmddp_K2_2phases} % and \ref{fig:PPM_K2_2phases} 
in the Supplementary Material. It is notable that partition-based models appear to identify fewer changepoints. For instance, the DRPM detected only 15 changepoints (4.3\% of the total time points), primarily at the beginning of the time series, and the LLDPM identified 24 changepoints (6.9\%). In contrast, the other alternative models,  namely the LDDP, WDDP and GMDDP, identified approximately 14\% of the changepoints, with 48, 50 and 51 changepoints, respectively, out of 349 time points. These observations instill confidence in the performance of the LLDPM, which thus might represent a reliable modeling option, particularly in real-world applications where the ground truth is unknown.

\subsection{Collaborative Perinatal Project data}\label{sec:cpp}

The CPP data were collected to investigate the potential effects of Dichlorodiphenyl-dichloroethylene (DDE) exposure on pregnancy outcomes. In addition to maternal serum DDE concentration, the dataset includes key variables for each enrolled pregnant woman, such as hospital of admission, smoking status, gestational age (in days), and newborn birth weight (in grams). It has been widely used in epidemiological research to examine the impact of prenatal environmental exposures on neonatal health, offering valuable insights for public health policies and maternal care strategies \citep[see, e.g.,][]{klebanoff2009collaborative}. The original study by \citet{longnecker2001association} investigated the relationship between DDE exposure and two primary outcomes: preterm delivery and small-for-gestational-age births.

\begin{figure}
    \centering
    \includegraphics%[width=0.6\textwidth] 
    [clip,trim=0cm 0cm 0.1cm 0.1cm,width=0.6\textwidth]{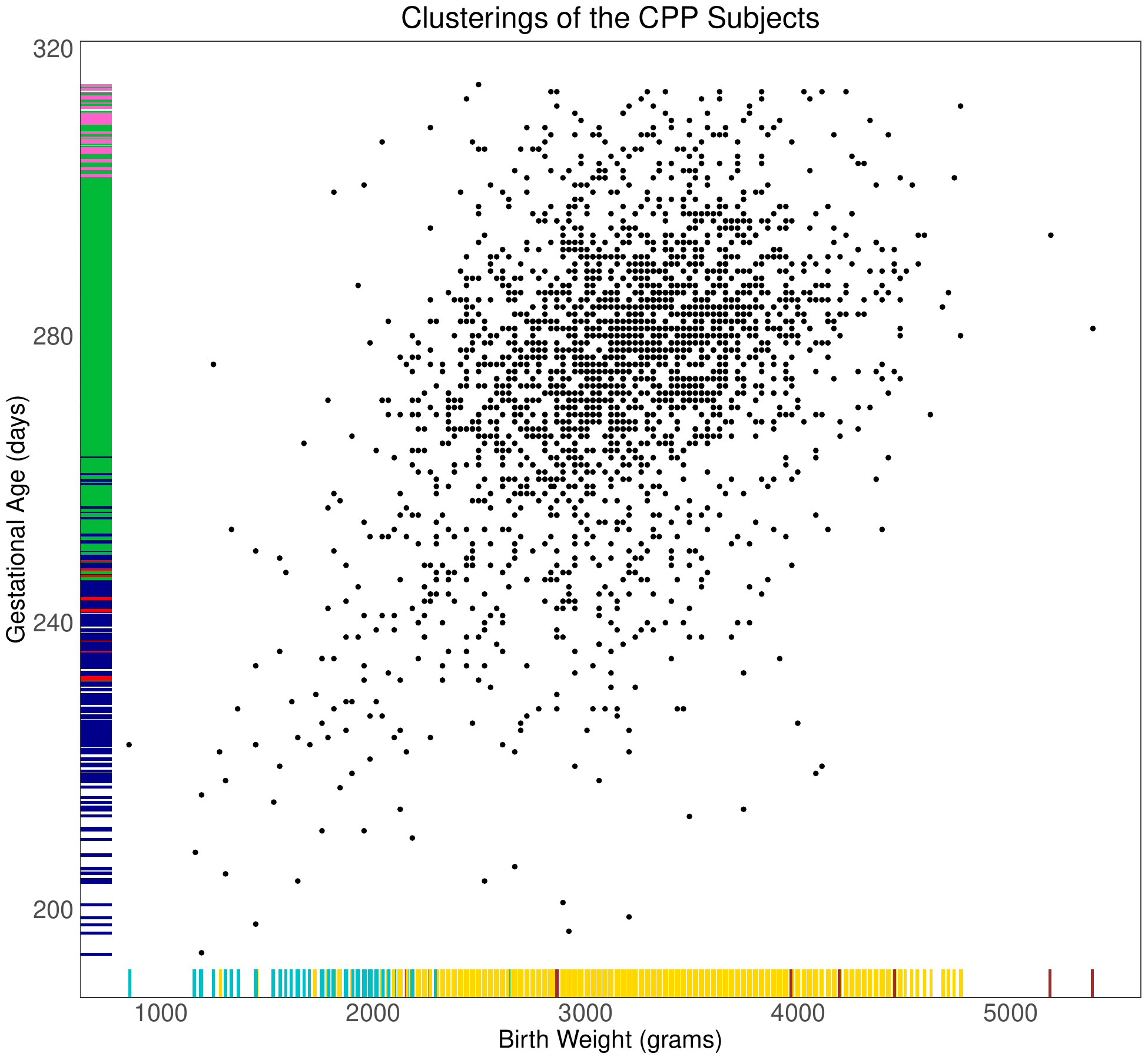}
    \caption{Section \ref{sec:cpp}, CPP data. Scatterplot of gestational age (days) and birth weight (grams), with colors indicating cluster assignment.}
    \label{fig:CPP_cluster2}
\end{figure}

In our analysis, we adopt the multiview framework outlined in Section \ref{sec:hierarcLDDP}, applying the LLDPM model to both gestational age and newborn birth weight. This approach treats the two measurements as distinct views of the same subjects. We focus on a subset of the CPP dataset, available in the \texttt{BNPmix} package \citep{corradin2021bnpmix}, which includes data from 2,312 women. The analysis is first conducted on the full dataset  and then stratified by hospital. In all cases, we assume a prior number of cluster equal to two and obtain $\theta=0.12$ when $n=2,312$. The prior probability of a change between the partitions of the two views is assumed $\eta \simiid \text{Beta}(0.1,0.9)$. For the distribution $P_0(\cdot)$ of the coefficients $\beta_{i,j}$, we assume a Normal distribution with a mean zero and variance $\varsigma^2 = 
1$, and we place an inverse-gamma prior on the variance parameter, $\tau^2 \sim \text{Inv-Gamma}
(2,1)$. We run our Gibbs sampling algorithm for 10,000 iterations, with the first half discarded as burn-in. 
Figure \ref{fig:CPP_cluster2} shows the scatterplot with the inferred view-specific clustering of the subjects 
displayed on the axis, estimated by minimizing the lower bound of the posterior expected VI  \citep{wade2018bayesian}.  The clusters identified from the first  view (gestational age) and the second view (birth weight)
exhibit patterns similar to those found in the analysis by \cite{dombowsky2025product}, whose model is specifically designed for multiview clustering. Gestational age can be divided into four clusters, each representing a different duration range of the pregnancy in days. 
For example, the pink cluster corresponds to longer pregnancies,  the blue cluster to shorter ones, and the red cluster to durations  around 240 days. Similarly, birth weight is categorized into three clusters: one  primarily includes newborns with  high birth weight, another captures those with lower birth weights, and the largest cluster corresponds to average-weight newborns. 
This discrepancy in the number and structure of clusters across the two views corresponds to what is defined as a changepoint when the framework is applied to model the evolution of partitions over time. Importantly, although the two inferred partitions differ, they reveal a common pattern: clusters linked to higher birth weights tend to align with those associated with longer gestational periods. This reflects a biologically plausible relationship: as gestational age increases, so does the expected birth weight.

To examine variability across hospitals, we perform the same analysis after stratifying the data by hospital. Table \ref{tab:hospital_results} reports, for each hospital, the sample size \(n\) and the value of \(\theta\), which is specified based on the assumption that the prior number of clusters is two for both views. Additionally, it summarizes the results of the LLDPM across hospitals, showing that the estimated probability \(\hat{\eta}\) of obtaining different partitions between the two views varies by hospital, ranging from 0.76 to 1. This variability underscores the model's ability to adapt when applied to subsets of different sizes.

\begin{table}[ht]
\centering
\begin{adjustbox}{max width=\textwidth}
\begin{tabular}{c|c|c|c|c|c|c|c|c|c|c|c|c}
Hospital  & 1      & 2      & 3      & 4      & 5      & 6      & 7      & 8      & 9      & 10     & 11     & 12  \\ 
\hline
$n$           & 481    & 124    & 150    & 77     & 205    & 154    & 141    & 141    & 117    & 384    & 151    & 188 \\ 
$\theta$ & 0.32 & 0.20 & 0.19 & 0.22 & 0.18 & 0.19 & 0.19 & 0.19 & 0.20 & 0.16 & 0.19 & 0.18\\
$\hat{\eta}$ & 1.00 & 0.87 & 1.00 & 0.77 & 1.00 & 0.92 & 0.99 & 1.00 & 1.00 & 1.00 & 1.00 & 1.00 \\  
\end{tabular}
\end{adjustbox}
\caption{Section \ref{sec:cpp}, CPP data stratified by hospital. For each of the 12 hospitals in the dataset, we report the sample size \(n\), the value of \(\theta\), and the estimated probability \(\hat{\eta}\) of a change between the latent partitions characterizing the two views: newborn birth weight and gestational age.}
\label{tab:hospital_results}
\end{table}

\section{Discussion}\label{sec:fine}

We introduced a new approach for modeling partitions with temporal dependence that relies on the definition of partition-based state equations in a state-space model, thus extending the dynamic linear model widely used in multivariate time series analysis. By characterizing the evolution of the local level equation across partitions and detecting changepoints, our model introduces dependence only when supported by the data. This is achieved by incorporating a mixture representation, reminiscent of the spike-and-slab prior commonly used in variable selection, to model the dependence between partitions. Here, a changepoint is defined as a shift in the partition of units between two consecutive time points. If this shift in cluster allocations coincides with a change in the process mean, our model can detect the changepoint in the traditional sense. However, if the process mean changes while the partition of units remains unchanged, the model will not identify a changepoint. This behavior reflects the model’s design: it is tailored to detect changes in clustering structure rather than in the underlying process itself. 
We have illustrated the model's performance using synthetic data, which highlighted its accuracy in recovering cluster structures and identifying changepoints. In an application to human gesture data, our model yields interpretable results by identifying most changepoints during periods of activity. 

Drawing upon the framework presented in Section \ref{sec:hierarcLDDP}, our model can be extended to analyze multivariate data across multiple sources, where dependence arise beyond temporal factors, e.g., varying experimental conditions. 
Importantly, also in this extended context, the random partitions maintain a marginal identical distribution, e.g. the distribution of a CRP. Although extending our algorithm for partition changepoint detection to multiview scenarios is not straightforward, an application to epidemiological data displays its effectiveness in a two-view clustering context. Another potential extension involves applications where multiple variables are observed across multiple subjects. Our model can be adapted to accommodate this added complexity, provided the data follow an experimental design in which changepoints are expected to be shared across subjects and the subjects are exchangeable. This structure does not apply to the human gesture dataset described in Section \ref{sec:gesturedata}, where, although data are available for multiple individuals, each subject recounts a different comic story. As a result, the time series reflect distinct underlying processes, and changepoints are not shared across subjects. A related example is offered by \citet{ricci2024bayesian}, who recently proposed a biclustering approach that groups subjects based on similarities in temporal trends observed during a task-based fMRI experiment.

Another exciting avenue for future research is the development of dependent random partition state-space models capable of modeling latent state equations that encode more complex dynamics. This extension could encompass higher-order time dependence, allowing for exploring more complex temporal relationships and patterns in data. In order to ensure smoother transitions over time, a hierarchical model formulation could shrink the base partition distribution $p^*(\cdot)$ to the partition estimated at the previous time or to a fixed ``anchor'' partition. However, care will be needed to ensure computational feasibility and interpretability of the latent variables $\gamma_t$. See, e.g., \cite{Paganin2021} or \cite{dahl2025dependent} for details on recently proposed anchored partitioned models. On a related note, while we have assumed that the changepoint selection indicators $\gamma_t$ are independent across time, they could instead be modeled as dependent on time-varying covariates, such as respiratory data or other measurements of expended effort in the analysis of human gesture data. 
Along similar lines, alternative strategies could be explored to enhance flexibility in modeling partition similarity over time. For instance, the changepoint indicator $\gamma_t$ could be modeled as a multinomial latent variable distinguishing between no change from the previous partition, a transition to a new partition, or a reversion to a previously observed partition. This approach could also build on the framework proposed by \citet{dahl2025dependent}. Alternatively, the mixture formulation in \eqref{eq:partition_state_equation}, describing the conditional distribution of $\pi_t$, could be recast as a linear combination between a base process $p^\ast(\cdot)$ and a more flexible dependent process in lieu of point mass at $\pi_{t-1}$.

Finally, we note that our model is designed to efficiently handle large $T$ through the introduction of the auxiliary changepoint variables $\gamma_t$. When no changepoint is detected, the partition remains unchanged across consecutive time points, allowing the algorithm to quickly proceed to the next iteration and significantly improve computational efficiently. Beyond changepoint detection, our framework opens the door to extensions in spatial and spatio-temporal domains. In purely spatial contexts, a multiview representation can be introduced, where spatial counterparts of the auxiliary changepoint variables detect changes in partitions across regions. This reconceptualizes the role of a changepoint as a shift in spatial structure, with clear relevance to biomedical tissue and transcriptomics data, where understanding spatial dependence among neighboring cells is critical. For spatio-temporal applications, the changepoint indicator $\gamma_{s,t}$ could be defined to vary simultaneously across time $t$ and spatial location $s$, with spatial covariates incorporated to capture region specific dynamics and interactions. Allowing these auxiliary changepoint variables to be modeled as spatio-temporally dependent would enable changepoint probabilities to evolve with both spatial and temporal covariates. Such developments would substantially expand the scope of our framework, offering novel tools for analyzing complex datasets where clustering structures evolve across both space and time.

\bibliographystyle{apalike}
\bibliography{main}

\section*{Supplementary Material}
The Supplementary Material contains the proofs of the results in Section \ref{sec:model}, additional details on posterior computation, an extended description of the changepoint detection procedure, and further information on the simulation studies and applications.

The Supplementary Material is organized as follows. In Section \ref{proof} we provide the proofs of the results in Section \ref{sec:model} of the main article. Section \ref{appendix:MCMC} provides additional details on posterior computations. In Section \ref{sec:changepoint} we provide more details on the changepoint detection procedure. Finally, Sections \ref{sim_data_appendix} and \ref{app:gesture} report further information on the simulation studies and the application, in Sections \ref{sec:sim} and \ref{sec:gesturedata} of the main article. 
\appendix
\section{Proofs}\label{proof}

\subsection{Proof of Proposition \ref{prop:ERI}}

\begin{proof}
     We let $c_{i,t}$, for $i=1,\ldots,n$ and $t=1,\ldots,T$, be a random variable taking values in $\{1,\ldots,|\pi_t|\}$, such that $c_{i,t}=j$ if $i\in C_{j,t}$, and $t_2 = t_1 + 1$ be two consecutive times. As observed in \cite{page2022dependent}, the ERI can be written as \begin{equation}\label{eq:ERIPROOF}    \varphi_{t_1,t_2}=\mathbb{E}[R(\pi_{t_1}, \pi_{t_2})] = 
    \binom{n}{2}^{-1}\sum_{1\leq i\leq j \leq n} \psi_{i,j},
    \end{equation} 
    where $\psi_{i,j} = P(c_{i,t_1} = c_{j,t_1} , c_{i,t_2} = c_{j,t_2}) + P(c_{i,t_1} \neq c_{j,t_1} , c_{i,t_2} \neq c_{j,t_2})$.
   Thanks to exchangeability, it is easy to verify that $\varphi_{t_1,t_2}=\psi_{1,2}$. 
   From the predictive distribution of a Gibbs-type prior, we obtain $P(c_{1,t_1} = c_{2,t_1}) = V_{2,1} (1-\sigma)$ and $P(c_{1,t_1} \neq c_{2,t_1}) = V_{2,2}$. Moreover, 
    $$
    P(c_{1,t_2} = c_{2,t_2} \mid c_{1,t_1} = c_{2,t_1}, \gamma_{t_2}) = \begin{cases}
        1 &\text{ if } \gamma_{t_2} = 0\\
        V_{2,1} (1-\sigma) &\text{ if }\gamma_{t_2} = 1,
    \end{cases}
    $$
    and
    $$
    P(c_{1,t_2} \neq c_{2,t_2} \mid c_{1,t_1} \neq c_{2,t_1}, \gamma_{t_2}) = \begin{cases}
        1 &\text{ if } \gamma_{t_2} = 0\\
        V_{2,2} &\text{ if } \gamma_{t_2} = 1.
    \end{cases}
    $$
    Then, we can calculate
    \begin{align}
    \begin{split}\label{eq:peq}
        P(c_{1,t_1} = c_{2,t_1}, c_{1,t_2} = c_{2,t_2}) =& P(c_{1,t_1} = c_{2,t_1}) \\[5pt]
        \times &
        \sum_{\gamma \in \{0,1\}} P(c_{1,t_2} = c_{2,t_2} \mid c_{1,t_1} = c_{2,t_1}, \gamma_{t_2}=\gamma)P(\gamma_{t_2}=\gamma)\\
        =& V_{2,1} (1-\sigma)(1-\eta) + \Big[V_{2,1} (1-\sigma)\Big]^2 \eta.
        \end{split}
    \end{align}
Similarly,
    \begin{align}
    \begin{split}\label{eq:pneq}
        P(c_{1,t_1} \neq c_{2,t_1}, c_{1,t_2} \neq c_{2,t_2}) =&P(c_{1,t_1} \neq c_{2,t_1})\\[5pt]
        \times&\sum_{\gamma \in \{0,1\}} P(c_{1,t_2} \neq c_{2,t_2} \mid c_{1,t_1} \neq c_{2,t_1}, \gamma_{t_2}=\gamma)P(\gamma_{t_2}=\gamma)\\
        =& V_{2,2} (1-\eta) + V_{2,2}^2 \ \eta.
        \end{split}
    \end{align}
    The result follows by summing \eqref{eq:peq} and \eqref{eq:pneq} and by recalling that $V_{1,1} = V_{2,1} (1-\sigma) + V_{2,2}$.
\end{proof}

\subsection{Proof of Proposition \ref{prop:ERIt1t2}}

\begin{proof}
     We generalize the proof of Proposition \ref{prop:ERI} under the more general assumption $t_1 < t_2$. We have
    \begin{align}
\begin{split}\label{eq:peq2}
        P(c_{1,t_1} = c_{2,t_1}, c_{1,t_2} = c_{2,t_2}) =& P(c_{1,t_1} = c_{2,t_1}) \\[5pt]
        \times &
        \sum_{\gamma \in \{0,1\}} P(c_{1,t_2} = c_{2,t_2} \mid c_{1,t_1} = c_{2,t_1}, \gamma_{t_2}=\gamma)P(\gamma_{t_2}=\gamma)\\
        =& V_{2,1} (1-\sigma)(1-\eta)^{t_2-t_1} + \Big[V_{2,1} (1-\sigma)\Big]^2 \left(1-(1-\eta)^{t_2-t_1}\right).
        \end{split}
    \end{align}
Similarly,
    \begin{align}
    \begin{split}\label{eq:pneq2}
        P(c_{1,t_1} \neq c_{2,t_1}, c_{1,t_2} \neq c_{2,t_2}) =&P(c_{1,t_1} \neq c_{2,t_1})\\[5pt]
        \times&\sum_{\gamma \in \{0,1\}} P(c_{1,t_2} \neq c_{2,t_2} \mid c_{1,t_1} \neq c_{2,t_1}, \gamma_{t_2}=\gamma)P(\gamma_{t_2}=\gamma)\\
        =& V_{2,2} (1-\eta)^{t_2-t_1} + V_{2,2}^2 \left(1-(1-\eta)^{t_2-t_1}\right).
        \end{split}
    \end{align}
    The result follows by recalling that $\psi_{i,j} = P(c_{i,t_1} = c_{j,t_1} , c_{i,t_2} = c_{j,t_2}) + P(c_{i,t_1} \neq c_{j,t_1} , c_{i,t_2} \neq c_{j,t_2})$ and by combining \eqref{eq:peq2} and \eqref{eq:pneq2} in \eqref{eq:ERIPROOF}.
\end{proof}

\subsection{Proof of Proposition \ref{prop:PSM_marginal}}
\begin{proof} 
By assumption, $\pi_1\sim p^*(\pi_1)$. For any $t=2,\ldots,T$, we denote by $p(\pi_t)$ the distribution of $\pi_t$, which can be obtained by marginalizing the distribution of $\bm{\pi}_{1:T}$ with respect to all the remaining $T-1$ components. We start by marginalizing it with respect to $\pi_1$, which gives:
\begin{align*}
p(\pi_t)&=\sum_{\substack{r=1\\r\neq t}}^{T}\sum_{\pi_r\in\mathscr{P}}
p^*(\pi_1)\prod_{s=2}^T\left[(1-\eta_s)\delta_{\pi_{s-1}}(\pi_s)+\eta_s\,p^*(\pi_{s})\right]\\
   &=\sum_{\substack{r=2\\r\neq t}}^{T}\sum_{\pi_r\in\mathscr{P}}\prod_{s=3}^{T} \left[(1-\eta_s)\delta_{\pi_{s-1}}(\pi_s)+\eta_s p^\ast(\pi_{s})\right]\\
    &\times\sum_{\pi_1\in\mathscr{P}}p^\ast(\pi_1)\left[(1-\eta_2)\delta_{\pi_{2}}(\pi_1)+\eta_2\,p^\ast(\pi_{2})\right]\\
    &=\sum_{\substack{r=2\\r\neq t}}^{T}\sum_{\pi_r\in\mathscr{P}}\prod_{s=3}^{T} \left[(1-\eta_s)\delta_{\pi_{s-1}}(\pi_s)+\eta_s\,p^\ast(\pi_{s})\right]\left[(1-\eta_2)p^\ast(\pi_2)+\eta_2\,p^\ast(\pi_{2})\right]\\
    &=\sum_{\substack{r=2\\r\neq t}}^{T}\sum_{\pi_r\in\mathscr{P}} p^\ast(\pi_{2}) \prod_{s=3}^{T} \left[(1-\eta_s)\delta_{\pi_{s-1}}(\pi_s)+\eta_s\,p^\ast(\pi_{s})\right].
\end{align*}
By iterating the same procedure for the next $t-2$ terms of the first sum, we get
\begin{align*}
p(\pi_t)&=p^\ast(\pi_{t}) \sum_{r=t+1}^{T}\sum_{\pi_r\in\mathscr{P}} \prod_{s=t+1}^{T} \left[(1-\eta_s)\delta_{\pi_{s-1}}(\pi_s)+\eta_s p^\ast(\pi_{s})\right]\\
&=p^*(\pi_t),
\end{align*}
where the last identity holds as $\prod_{s=t+1}^{T} \left[(1-\eta_s)\delta_{\pi_{s-1}}(\pi_s)+\eta_s p^\ast(\pi_{s})\right]$ is the conditional distribution of $\bm{\pi}_{(t+1):T}$ given $\pi_t$, which thus sums up to one. Then, $\pi_t$ is marginally distributed as $p^\ast(\pi_t)$ for every $t=1,\ldots,T$.
\end{proof}

\subsection{Proof of the multiview clustering representation}\label{sm:multiview}

The joint distribution of $(\tilde\pi,\pi_1,\pi_2)$ and $(\gamma_1,\gamma_2)$ can be written as
\begin{equation*}
\begin{aligned} p(\bm{\pi}_{1:2},\tilde\pi,\tilde{\gamma}_1,\tilde{\gamma}_2)&=p(\tilde\pi,\pi_1,\pi_2\mid \gamma_1,\gamma_2)\, p(\gamma_1,\gamma_2)\\
    &=p(\tilde\pi)\, p(\pi_1\mid \tilde\pi,\gamma_1,\gamma_2)\, p(\pi_2\mid \tilde\pi,\gamma_1,\gamma_2)\, p(\gamma_1)\, p(\gamma_2)\\
    &=p^*(\tilde\pi)\, \left\{\prod_{t=1}^2 [(1-  \tilde\gamma_t)\delta_{\tilde\pi}(\pi_t)+\tilde\gamma_t\, p^*(\pi_t)]\,
    \tilde\eta^{\tilde\gamma_t}(1-\tilde\eta)^{1-\tilde\gamma_t}\right\}.
\end{aligned}
\end{equation*}
The distribution of $(\pi_1,\pi_2)$ is obtained by marginalizing the previous expression with respect to $(\gamma_1,\gamma_2)$ and $\tilde\pi$,
\begin{equation}
\begin{aligned} 
    p(\bm{\pi}_{1:2})&=\sum_{\tilde\pi\in\mathscr{P}}p^*(\tilde\pi)\prod_{t=1}^2 [(1-\eta_t)\delta_{\tilde\pi}(\pi_t)+\eta_t\, p^*(\pi_t)]\\
    &=p^*(\pi_1)\left[(1-\tilde{\eta})\, \delta_{\pi_1}(\pi_2)+\tilde{\eta}\, p^*(\pi_2)\right],
\label{eq:alternative_model_joint}
\end{aligned}
\end{equation}
where $\mathscr{P}$ indicates the space of all partitions 
and $\tilde{\eta}=1-(1-\eta_1)(1-\eta_2).$
By comparing \eqref{eq:joint_pi2} and \eqref{eq:alternative_model_joint}, we can notice that the two distributions coincide provided that $\tilde\eta=\eta$, which is achieved if 
$(1-\eta_1)(1-\eta_2)=(1-\eta)$. If we further make the assumption that $\eta_1=\eta_2$, then  $\eta_1=1-\sqrt{1-\eta}$, such that $\gamma_t \simiid \text{Bern}(1-\sqrt{1-\eta})$ in \eqref{eq:alternative_model_2}, $t=1, 2$.

\section{Further details on posterior computations}\label{appendix:MCMC}

\subsection{Gibbs sampler}\label{sec:Gibbs}
We briefly describe the updates of the model parameters at a generic iteration. At each iteration the random parameters involved in the MCMC scheme are the partitions $\bm{\pi}_{1:T}$, the auxiliary variable $\bm{\gamma}_{2:T}$, the dependence parameter $\eta_2,\ldots,\eta_T$ and the local level parameters $\bm{\beta}_1,\ldots,\bm{\beta}_T$. For simplicity, we do not introduce explicit notation to indicate at which iteration random quantities have been updated. Instead, the full conditional distributions for each update should be understood as being conditional on the most up-to-date values of the random quantities involved in the algorithm.
The algorithm is presented for the case $p^*(\cdot)=p_{\text{CRP}}(\cdot)$. Extending it to other base partition distributions is conceptually straightforward. 
 \begin{itemize}
     \item[1)] \textit{ Update of  $(\pi_t,\gamma_t)$}.
     For each $t = 1, \ldots, T$, we follow a two-step procedure, detailed below for $t = 2, \ldots, T - 1$, with straightforward adjustments for the boundary cases $t = 1$ and $t = T$.  
     \begin{itemize}
         \item[1.1)] We update $\pi_t$ from its full conditional distribution, computed after marginalizing $\gamma_t$ out. Namely  
         \begin{equation}\label{eq:full_pit}
         \begin{aligned}
         p(\pi_t\mid \ldots)&\propto \Big[ (1-\eta_t) \, \delta_{\pi_{t-1}}(\pi_t) + \eta_t \, p_{\text{CRP}} \, (\pi_{t})\Big]\\
       &\times \Big[(1-\gamma_{t+1})\,  \delta_{\pi_{t+1}}(\pi_t) + \gamma_{t+1} \,   p_{\text{CRP}} (\pi_{t+1}) \Big] \, p(\bm{Y}_t\mid \pi_t).
     \end{aligned}
     \end{equation}
The form of the full conditional in \eqref{eq:full_pit} highlights that the update of $\pi_t$ depends also on $\gamma_{t+1}$, the changepoint variable at the subsequent time step. We then distinguish between two cases:
    \begin{itemize}
    \item[1.1a)] If $\gamma_{t+1}=1$, indicating a changepoint at time $t+1$, then no information about the distribution of $\pi_t$ is borrowed from the subsequent time step. Thus, the full conditional of $\pi_t$ boils down to the following mixture: \\
    \begin{equation*}
       p(\pi_t \mid \gamma_{t+1}=1, \ldots)
        \propto (1-\eta_t)\,  p(\bm{Y}_t\mid \pi_{t-1}) \, \delta_{\pi_{t-1}}(\pi_t) + \eta_t \, p_{\text{CRP}}(\pi_{t})\,  p(\bm{Y}_t\mid \pi_t).
    \end{equation*}
   That is, $\pi_t$ coincides with $\pi_{t-1}$ with probability proportional to $(1-\eta_t)p(\bm{Y}_t\mid \pi_{t-1})$. Alternatively, it is generated as a random draw of a new partition from a random partition model with distribution proportional to $p_{\text{CRP}} (\pi_{t}) p(\bm{Y}_t\mid \pi_t)$. The probability of choosing this mixture component is   proportional to $\eta_ t g_t$, where
   \begin{equation}\label{eq:marg_lik}
        g_t=\sum_{\pi_t \in\mathscr{P}}{p_{\text{CRP}} (\pi_{t}) p(\bm{Y}_t\mid \pi_t)}
   \end{equation}
   is the marginal likelihood specific to time $t$. Later in the section we discuss how to evaluate $g_t$ and how to sample from a distribution proportional to $p_{\text{CRP}} (\pi_{t}) p(\bm{Y}_t\mid \pi_t)$. 
    \item[1.1b)] If $\gamma_{t+1} = 0$, indicating there is not a changepoint at time $t+1$, then the full conditional distribution of $\pi_t$ is degenerate at $\pi_{t+1}$. Thus we simply set $\pi_{t}$ equal to the current value of $\pi_{t+1}$.
    \end{itemize}
  \item[1.2)] We update the changepoint variable $\gamma_t$ from its full conditional distribution, thus conditioning on the value of $\pi_t$ as updated in step 1.1. This is equivalent with sampling a Bernoulli random variable with probability of success
  \begin{equation*}
     \Pr(\gamma_t=1\mid \ldots)=\frac{\eta_t p_{\CRP}(\pi_t)}{\eta_t p_{\CRP}(\pi_t)+(1-\eta_t)\delta_{\pi_{t-1}}(\pi_t)}.
  \end{equation*}
We observe that, if $\pi_t \neq \pi_{t-1}$, then $\gamma_t = 1$ almost surely.
     \end{itemize}
         
     \item[2)] \textit{Update of $\eta_t$}. For any $t=1,\ldots,T$, we update $\eta_t$ from its full conditional distribution, which is given by 
     \begin{equation*}
         \eta_t\mid \ldots \simind \text{Beta}(a+\gamma_t,b+1-\gamma_t).
     \end{equation*}
    
     \item[3)] \textit{Reshuffling step}. Conditionally on $\bm{\gamma}_{2:T}$, we update the partitions $\bm{\pi}_{1:T}$ using a sampling importance resampling step, as described in Section \ref{sec:reshuffling}.
 \end{itemize}
Step 1.1a highlights the computational challenges that arise when implementing our PSM. Although the partition-based state equation \eqref{eq:partition_state_equation} has the same mixture representation that characterizes tractable priors, such as the spike-and-slab prior commonly used for variable selection, working with random partitions makes computations more complex. The time-specific marginal likelihood \eqref{eq:marg_lik} is estimated via Monte Carlo, a step that, conveniently, must be performed only once, prior to running the Gibbs sampler. Section \ref{sec:burnin}  provides further details on how this step can be made efficient. Step 1.1a also involves sampling from a distribution proportional to $p_{\text{CRP}}(\pi_t)p(\bm{Y}_t\mid \pi_t)$, for every $t=1,\ldots,T$. To this end, we propose implementing an auxiliary MCMC run prior to initializing the primary MCMC algorithm for model fitting. In this preliminary run, we fit an independent CRP model at each time point, thereby obtaining a catalog of partitions that enables updating the individual clustering labels. We then save all the partitions generated throughout the MCMC iterations, creating, for each \(t=1,\ldots,T\), a catalog \(\mathscr{S}_t\) of realizations from the posterior distribution of \(\pi_t\) given \(\bm{Y}_t\). 
The size of these catalogs can be arbitrarily large. To sample a new partition from a distribution proportional to \(p_{\text{CRP}} (\pi_{t}) p(\bm{Y}_t\mid \pi_t)\), we refer to the relevant catalog of realized partitions from the auxiliary MCMC run, randomly selecting a new partition from \(\mathscr{S}_t\), for \(t=1, \ldots, T\). This strategy is crucial for making the resulting MCMC algorithm efficient and capable of handling observations recorded over a large number of time points \(T\). For instance, this is the case in the application presented in Section \ref{sec:gesturedata}, where \(T=349\). The described MCMC algorithm was implemented using the \texttt{R} software \citep{R} combined with the \texttt{C++} programming language via \texttt{RcppArmadillo} \citep{RcppArmadillo}. For a dataset consisting of \(100\) time series observed over \(100\) time points, the average computational time required for \(1,000\) iterations of our Gibbs sampling algorithm is approximately \(951.1\) seconds on a machine with an i7 processor and 16GB of RAM. Generating catalogs of \(5,000\) partitions in the pre-sampling step requires only 40 seconds for the same dataset on the same machine using 5 cores. The pre-sampling step can be parallelized across multiple cores, as data from different time points are dealt with independently.
Generalizing the presented algorithm to the multiview setting requires additional considerations.
Specifically, the algorithm adapts naturally to the two-view case with minimal adjustments, as shown in the analysis of Section \ref{sec:cpp}, essentially due to the equivalence of the two representations of the joint distribution of $\bm{\pi}_{1:2}$ provided in \eqref{eq:joint_pi2} and \eqref{eq:alternative_model_1}--\eqref{eq:alternative_model_2}. In contrast, extending the algorithm to more than two views requires designing a dedicated algorithm that explicitly updates the common parent partition $\tilde{\pi}$.

\subsection{Estimating the marginal likelihood $g_t$}\label{sec:burnin}
For any $t=1,\ldots,T$, the marginal likelihood $g_t$, as defined in \eqref{eq:marg_lik}, can be estimated via Monte Carlo. We observe that  
\begin{equation*}
        g_t=\sum_{\pi_t \in\mathscr{P}}{p_{\text{CRP}} (\pi_{t}) p(\bm{Y}_t\mid \pi_t)}=\mathbb{E}\left[p(\bm{Y}_t\mid \pi)\right],
   \end{equation*}
   where the expected value is taken with respect to a random partition $\pi\sim p_{\text{CRP}}(\pi)$. This step is made efficient based on three considerations. 
   \begin{itemize}
   \item[i)] Each $g_t$ is estimated only once, before running the Gibbs sampling described in Section \ref{sec:post}.
   \item[ii)] The same sample of partitions generated from $p_{\text{CRP}}(\cdot)$ can be used to estimate $g_t$ for all $t=1,\ldots,T$.
   \item[iii)] If the kernel $p(y_{i,t}\mid \beta_{i,t})$ in \eqref{eq:likelihood}, and the base measure $P_0(\cdot)$ are specified in a convenient form, the function $p(\bm{Y}_t\mid \pi_t)$ is available in closed-form and easy to evaluate. For example, when both are assumed normal (specifically $y_{i,t}\mid \beta_{i,t}\sim \text{N}(\beta_{i,t},\tau^2)$ and $P_0(\cdot)$ Normal with mean $\mu$ and variance $\varsigma^2$), as implemented in Sections \ref{sec:sim} and \ref{sec:gesturedata}, then 
   \begin{align*}
    \log p(\bm{Y}_t|\pi_t) = \sum_{j=1}^{|\pi_t|} -& \frac{n_j}{2} \log(2\pi) - n_j \log \tau - \log \varsigma + \frac{1}{2} \log \frac{\tau^2 \varsigma^2}{n_j \varsigma^2 + \tau^2}\\
    -&\frac{1}{2} \sum_{i=1}^{n_j} y_{i,t}^2 - \frac{\mu^2}{2 \varsigma^2} + \frac{\tau^2 \varsigma^2}{2(n_j \varsigma^2 + \tau^2)} \left(\frac{\mu}{\varsigma^2}+\frac{\sum_{i=1}^{n_j} y_{i,t}}{\tau^2}\right).
\end{align*}
   \end{itemize}

\subsection{Reshuffling step}\label{sec:reshuffling}

Updating the partitions $\bm{\pi}_{1:T}$, conditionally on $\bm{\gamma}_{2:T}$, at each iteration of the Gibbs sampler helps improving the mixing of the algorithm. 
We start by observing that the values taken by $\bm{\gamma}_{2:T}$ determine a partition of the partitions $\pi_{1:T}$ into $k_T\leq T$ blocks. For example, if $\gamma_2=\gamma_3=0$ and $\gamma_4=1$, then the first block consists of the three identical partitions $\{\pi_1,\pi_2,\pi_3\}$. Within each block, all partitions are identical. Given the finiteness of $\mathscr{P}$, there may be ties also among partitions belonging to different blocks. We introduce the notation $\{\pi_1^*,\ldots,\pi_{k_{T}}^*\}$ to indicate the $k_{T}$ partitions corresponding to the $k_T$ blocks.
The reshuffling step consists in updating the values $\pi_j^*$, for $j=1,\ldots,k_T$, from the corresponding full conditional distributions. To streamline notation, we define $\mathcal{D}_j$ as the set of indices in $\{1,\ldots,T\}$ identifying the partitions belonging to the $j$th block of the partition of $\bm{\pi}_{1:T}$ determined by $\bm{\gamma}_{2:T}$. Referring to the previous example, we have $\mathcal{D}_1 = \{1,2,3\}$.
The full conditional of $\pi_j^*$ takes the form
\begin{equation}\label{eq:fc_pistar}
    p(\pi_j^*\mid \ldots)\propto p_{\text{CRP}}(\pi_j^\ast) \prod_{l\in \mathcal{D}_j}  p(\bm{Y}_l\mid \pi_j^\ast).
\end{equation}
Recall that $\mathscr{S}_{t_j}$ is a catalog of realizations from the posterior distribution of $\pi_{t_j}$ given $\bm{Y}_t$. We note that these are generated in a preliminary step and are therefore treated as given at the start of this step. 
If $|\mathcal{D}_j|=1$, say $\mathcal{D}_j=\{t_j\}$,
for some time $t_j$, 
then the problem of sampling from \eqref{eq:fc_pistar} can be tackled simply by picking an element at random from the list $\mathscr{S}_{t_j}$. 
If $|\mathcal{D}_j|>1$, say $\mathcal{D}_j=\{t_{j},\ldots,t_{j}+|\mathcal{D}_j|-1\}$, for some time $t_j$, then we propose a sampling importance resampling (SIR) step to simulate from \eqref{eq:fc_pistar}. The SIR algorithm proceeds by drawing a sample from an importance sampling function $g(\cdot)$, referred to as the envelope, which we define as a categorical distribution with categories  $\mathscr{S}^{(j)}=\{\pi_1^{(j)},\ldots,\pi_{M_j}^{(j)}\}$, defined by pooling the catalogs $\mathscr{S}_t$, for all $t\in\mathcal{D}_j$. Given that ties may be present both within and across catalogs, each category $\pi_i^{(j)}$, for $i=1,\ldots,M_j$, will be associated with a frequency $m_i^{(j)}$. Thus we can write $g(\pi_i^{(j)})\propto m_{i}^{(j)}$. The SIR algorithm then proceeds as follows.
\begin{enumerate}
    \item[i.] For every $j$,
    we sample at random $m$ draws, with replacement, from $\mathscr{S}^{(j)}$, say $\mathscr{M}^{(j)}=\{\tilde\pi_{1}^{(j)},\ldots,\tilde\pi_{m}^{(j)}\}$. 
    \item[ii.] For each $i=1,\ldots,m$, we assign a weight $w_i^{(j)}$ to each drawn partition $\tilde\pi_i^{(j)}$, computed by evaluating at $\pi_{i}^{(j)}$ the ratio between the target distribution in \eqref{eq:fc_pistar} and the envelope $g(\cdot)$. Namely,
\begin{align*}
w_i^{(j)}&=\frac{p(\tilde\pi_{i}^{(j)}\mid \{Y_t\}_{t \in D_j})}{g(\tilde\pi_i^{(j)})}\\
&\propto \frac{p_{\text{CRP}}(\tilde\pi_i^{(j)}) \prod_{l\in \mathcal{D}_j}  p(\bm{Y}_l\mid \tilde\pi_i^{(j)})}{m_i^{(j)}}.
\end{align*} 
    \item[iii.] We sample one value for $\pi_j^\ast$ out of the candidates in $\mathscr{M}^{(j)}$, with probabilities proportional to $\{w_1^{(j)}, \ldots, w_{m}^{(j)}\}$.
\end{enumerate}
Following standard results \citep[see, e.g.,][]{Ska03}, this yields a valid sampling strategy, provided that $m$ is sufficiently large.
The proposed SIR method offers an efficient and practical solution. Still, alternative approaches, such as Metropolis-Hastings, can also be employed for this reshuffling step, either independently or in combination with SIR.

\section{More details on changepoint detection}\label{sec:changepoint}

For each time point $t=2,\ldots,T$, 
 the detection of a changepoint at $t$ is related to the probability of a changepoint at times $t-1$ and $t+1$.  That is, the null hypothesis $H_{0, t}$, i.e. no changepoint at time $t$, is assessed with respect to the related decisions at times $t-1$ and $t+1$.
Considering this set of dependent hypotheses together is crucial, since, for example, a false changepoint detection at time $t-1$ may induce a false changepoint detection at time $t$, even if the null hypothesis is true at both times. We let  $d_t$ represent the decision at time $t$, i.e., $d_t = 1$ if the $t$-th hypothesis is rejected and $d_t = 0$ if it is not. Similarly, $r_t$ denotes the truth at time $t$, i.e., $r_t=0$ %$r_t=1$ 
if  $H_{0,t}$ is true, $r_t=1$ otherwise. 
We also introduce the notation $\bm{d}=(d_2, \ldots, d_T)$, $\bm{r}=(r_2, \ldots, r_T)$. We define the true positive rate (TPR) as the ratio between the number of times where the decision correctly identifies a changepoint, and the number of positive decisions. That is 
$$\text{TPR}(\boldsymbol{d}, \boldsymbol{r})=\frac{1}{D}\sum_{t=2}^T d_t r_t,
$$
where $D=\sum_{t=2}^T d_t$. 
We consider the compound loss function 
$$
L(\boldsymbol{d}, \boldsymbol{r})=-  \text{TPR}(\boldsymbol{d}, \boldsymbol{r}) + \kappa \, \text{ER}(\boldsymbol{d}, \boldsymbol{r}),
$$
where $\kappa$ is a positive constant, and, in order to penalize false detections at each time $t$, 
the error rate (ER) is defined as the ratio between the total number of false detections in the set of time points $\mathcal{G}_t$ and the number of positive decisions $D$. That is,
\begin{equation}
    \begin{aligned}
        \text{ER}(\bm{d},\bm{r}) &= \frac{1}{D}\left\{\sum_{t=3}^{T} d_{t-1} (1-r_{t-1}) + \sum_{t=2}^T d_t (1-r_t) +  \sum_{t=2}^{T-1} d_{t+1} (1-r_{t+1})\right\}\\
        &= \frac{1}{D}\left\{  d_{2} (1-r_{2}) + 3 \sum_{t=2}^{T}  d_t  (1-r_t) +   d_T  (1-r_T)\right\}.
        \label{eq:ER}
    \end{aligned}
\end{equation}
 While the true changepoint times $t$, where $r_t = 1$, are unknown, they can be estimated using the corresponding changepoint indicators $\gamma_t$. Therefore, we seek to minimize the posterior expected loss $\mathbb{E}[L(\bm{d}, \bm{\gamma}) \mid \bm{Y}_{1:T}]$ with respect to $\bm{d}$, where the expectation is taken over the posterior distribution of $\bm{\gamma}$, which is compact notation for $\bm{\gamma}_{2:T}$. This leads to the following optimization problem for identifying the changepoints:
\begin{align*}
\min_{\boldsymbol{d}} \, \mathbb{E}[L(\boldsymbol{d}, \bm{\gamma})\mid\bm{Y}_{1:T}]&= \min_{\boldsymbol{d}} \left\{-  \mathbb{E}[\text{TPR}(\boldsymbol{d}, \bm{\gamma})\mid \bm{Y}_{1:T}] + \kappa \,\mathbb{E}[\text{ER}(\boldsymbol{d}, \bm{\gamma}_{2:T})\mid \bm{Y}_{1:T}]\right\}\\
&=\min_{\boldsymbol{d}} \left\{-  \text{TPR}(\boldsymbol{d}, \boldsymbol{\text{PPC}}) + \kappa\, \text{ER}(\boldsymbol{d}, \bm{\text{PPC}})\right\},
\end{align*}
where $\boldsymbol{\text{PPC}}$ is the vector of posterior probabilities of a changepoint, $\text{PPC}_t$, $t=2, \ldots, T$. Following a similar approach to Theorem 1 in \citet{muller2004optimal}, it can be shown that the optimal decision rule corresponds to setting a threshold on the posterior probabilities $\text{PPC}_t$. 
Additionally, because the expression for the error rate in equation \eqref{eq:ER} includes the term \( 3 \sum_{t=2}^{T} d_t (1-r_t) \), the resulting non-marginal Bayesian FDR, denoted as \(\text{BFDR}_{\text{nm}}\), can be controlled such that \(\text{BFDR}_{\text{nm}}(h) = 3\,  \text{BFDR}_{\text{m}}(h)\), where \(\text{BFDR}_{\text{m}}\) is defined in \eqref{eq:FDR}. Thus, it is possible to control the non-marginal \(\text{BFDR}_{\text{nm}}\) by computing the commonly used marginal Bayesian FDR and considering a more stringent level, say $\zeta/3$, for the identification of the optimal threshold on the $\text{PPC}_{t}$'s. This correction becomes especially relevant when dealing with autoregressive data, which are characterized by higher structural dependence.

\section{More on the simulation studies of Section \ref{sec:sim}}\label{sim_data_appendix}
We present a description of the metrics used in the simulation studies, and additional plots related to the simulated data in Section \ref{sec:sim}.

\subsection{Metrics}\label{sm:metrics}
Before reviewing the definitions of specificity, accuracy, recall, precision, F1 score, and AUC, we introduce the following terms:
\begin{itemize}
    \item True Positives (TP): changepoints correctly detected.  
    \item False Positives (FP): detected changepoints that do not correspond to any true changepoint.  
    \item False Negatives (FN): true changepoints that were not detected.  
    \item True Negatives (TN): non-changepoint times correctly identified as such.  
\end{itemize}
The specificity measures the proportion of true negatives correctly identified, that is
\[
\text{Specificity} = \frac{\text{TN}}{\text{TN} + \text{FP}}.
\]
The accuracy quantifies the overall correctness of the changepoint detection, that is
\[
\text{Accuracy} = \frac{\text{TP} + \text{TN}}{\text{TP} + \text{TN} + \text{FP} + \text{FN}}.
\]
The recall measures the proportion of actual changepoints that were correctly detected, that is
\[
\text{Recall} = \frac{\text{TP}}{\text{TP} + \text{FN}}.
\]
The precision assesses the proportion of detected changepoints that are actually correct, that is
\[
\text{Precision} = \frac{\text{TP}}{\text{TP} + \text{FP}}
\]
The F1 Score is the harmonic mean of precision and recall, that is
\[
\text{F1 Score} = 2 \times \frac{\text{Precision} \times \text{Recall}}{\text{Precision} + \text{Recall}}
\]
The Area Under the Curve (AUC) measures the overall performance of a changepoint detection method by evaluating the trade-off between true positive rate (recall) and false positive rate.

Table \ref{tab:thresholds} shows the values of these metrics for different competing methods across different thresholds (see Section \ref{sec:sim} for more details).

\begin{table}[htbp]
\centering
\caption{Optimized Values for Different Thresholds}
%\begin{adjustbox}{max width=\textwidth}
\begin{tabular}{@{}llcccccccc@{}}
\toprule
\multirow{2}{*}{Measure} & \multirow{2}{*}{Method} & \multicolumn{6}{c}{Threshold} \\ \cmidrule(l){3-8} 
                         &                         & 0.75  & 0.8   & 0.85  & 0.9   & 0.95  & 1.0   \\ \midrule
\multirow{4}{*}{Specificity} 
                         & DRPM                    & 0.99 & 0.96 & 0.91 & 0.55 & 0.55 & 0.21 \\
                         & LDDP                    & 0.96 & 0.93 & 0.67 & 0.12 & 0.01 & 0.01 \\
                         & WDDP                    & 0.68 & 0.54 & 0.28 & 0.02 & 0.03 & 0.01 \\
                         & GMDDP                   & 1     & 1     & 1     & 1     & 1     & 1     \\ \midrule
\multirow{4}{*}{Accuracy} 
                         & DRPM                    & 0.93 & 0.93 & 0.89 & 0.58 & 0.59 & 0.27 \\
                         & LDDP                    & 0.94 & 0.94 & 0.69 & 0.19 & 0.09 & 0.09 \\
                         & WDDP                    & 0.69 & 0.57 & 0.34 & 0.10 & 0.11 & 0.09 \\
                         & GMDDP                   & 0.95 & 1     & 1     & 1     & 1     & 1     \\ \midrule
\multirow{4}{*}{Recall} 
                         & DRPM                    & 0.28 & 0.51 & 0.64 & 0.95 & 1     & 1     \\
                         & LDDP                    & 0.65 & 1     & 1     & 1     & 1     & 1     \\
                         & WDDP                    & 0.82 & 0.89 & 1     & 1     & 1     & 1     \\
                         & GMDDP                   & 0.37 & 0.99 & 1     & 1     & 1     & 1     \\ \midrule
\multirow{4}{*}{Precision} 
                         & DRPM                    & 0.78 & 0.72 & 0.60 & 0.38 & 0.35 & 0.14 \\
                         & LDDP                    & 0.91 & 0.86 & 0.49 & 0.09 & 0.08 & 0.08 \\
                         & WDDP                    & 0.18 & 0.15 & 0.11 & 0.08 & 0.08 & 0.08 \\
                         & GMDDP                   & 1     & 1     & 1     & 1     & 1     & 1     \\ \midrule
\multirow{4}{*}{F1}      
                         & DRPM                    & 0.39 & 0.49 & 0.55 & 0.47 & 0.45 & 0.22 \\
                         & LDDP                    & 0.70 & 0.88 & 0.57 & 0.17 & 0.15 & 0.15 \\
                         & WDDP                    & 0.30 & 0.25 & 0.20 & 0.15 & 0.15 & 0.15 \\
                         & GMDDP                   & 0.55 & 1 & 1     & 1     & 1     & 1     \\ \midrule
\multirow{4}{*}{AUC}     
                         & DRPM                    & 0.59 & 0.69 & 0.78 & 0.75 & 0.78 & 0.60 \\
                         & LDDP                    & 0.76 & 0.94 & 0.73 & 0.56 & 0.51 & 0.51 \\
                         & WDDP                    & 0.75 & 0.72 & 0.64 & 0.51 & 0.52 & 0.51 \\
                         & GMDDP                   & 0.69 & 1 & 1     & 1     & 1     & 1     \\ \bottomrule
\end{tabular}
%\end{adjustbox}
\label{tab:thresholds}
\end{table}

\subsection{Plots}\label{sm:plots}

\begin{figure}[h!]
    \centering
    \includegraphics[clip,trim=0cm 0cm 0cm 0.15cm,width=0.85\textwidth]{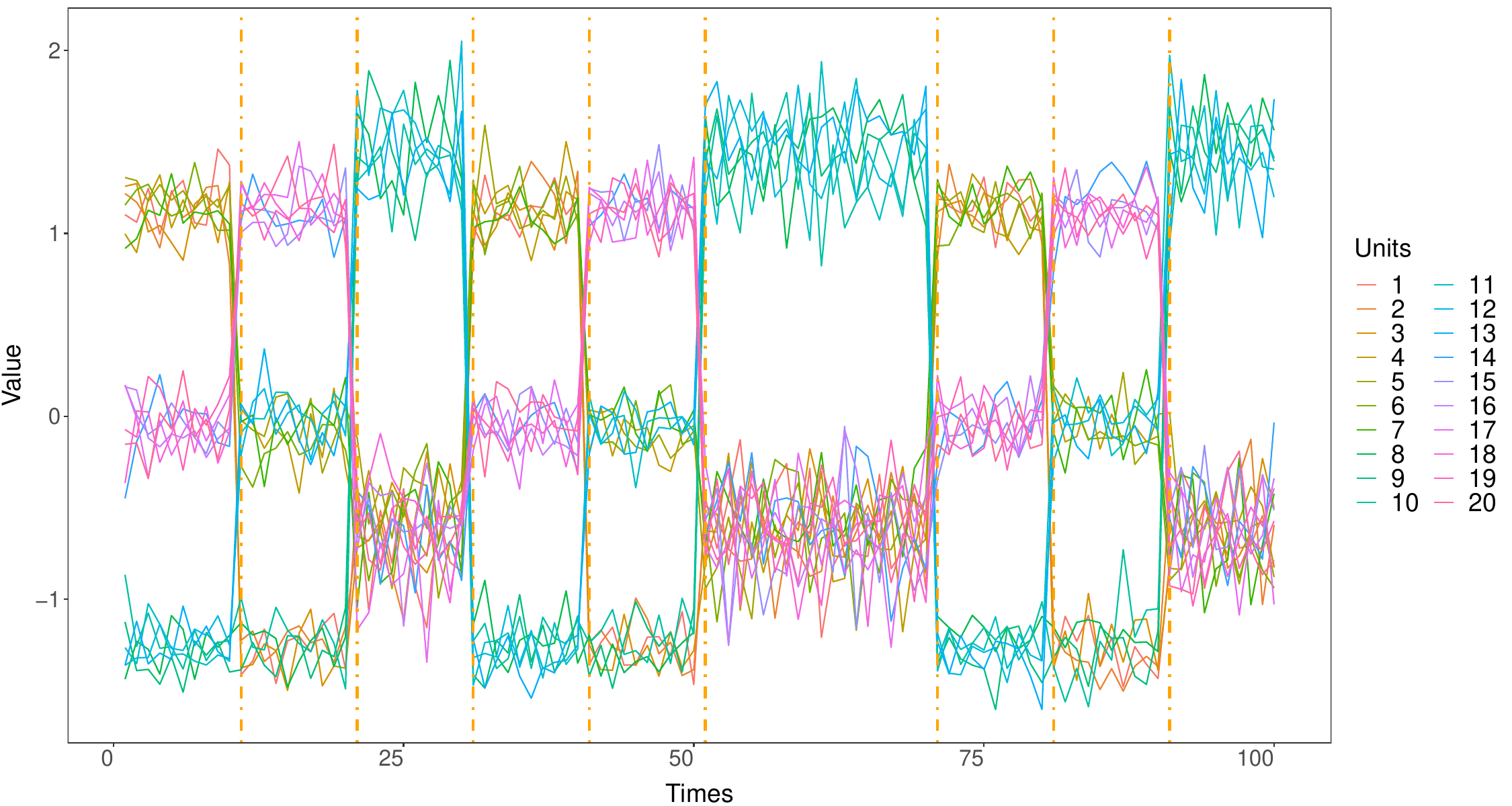}
    \caption{Example of simulated independent dataset, with $n=20$ subjects and $T=100$ time points. The orange vertical lines correspond to changepoints. See Section \ref{sec:indep} for details.}
    \label{fig:indep_data}
\end{figure}

\begin{figure}[h!]
\centering
 \includegraphics[clip,trim=0cm 0cm 0cm 0.15cm,width=0.55\textwidth]{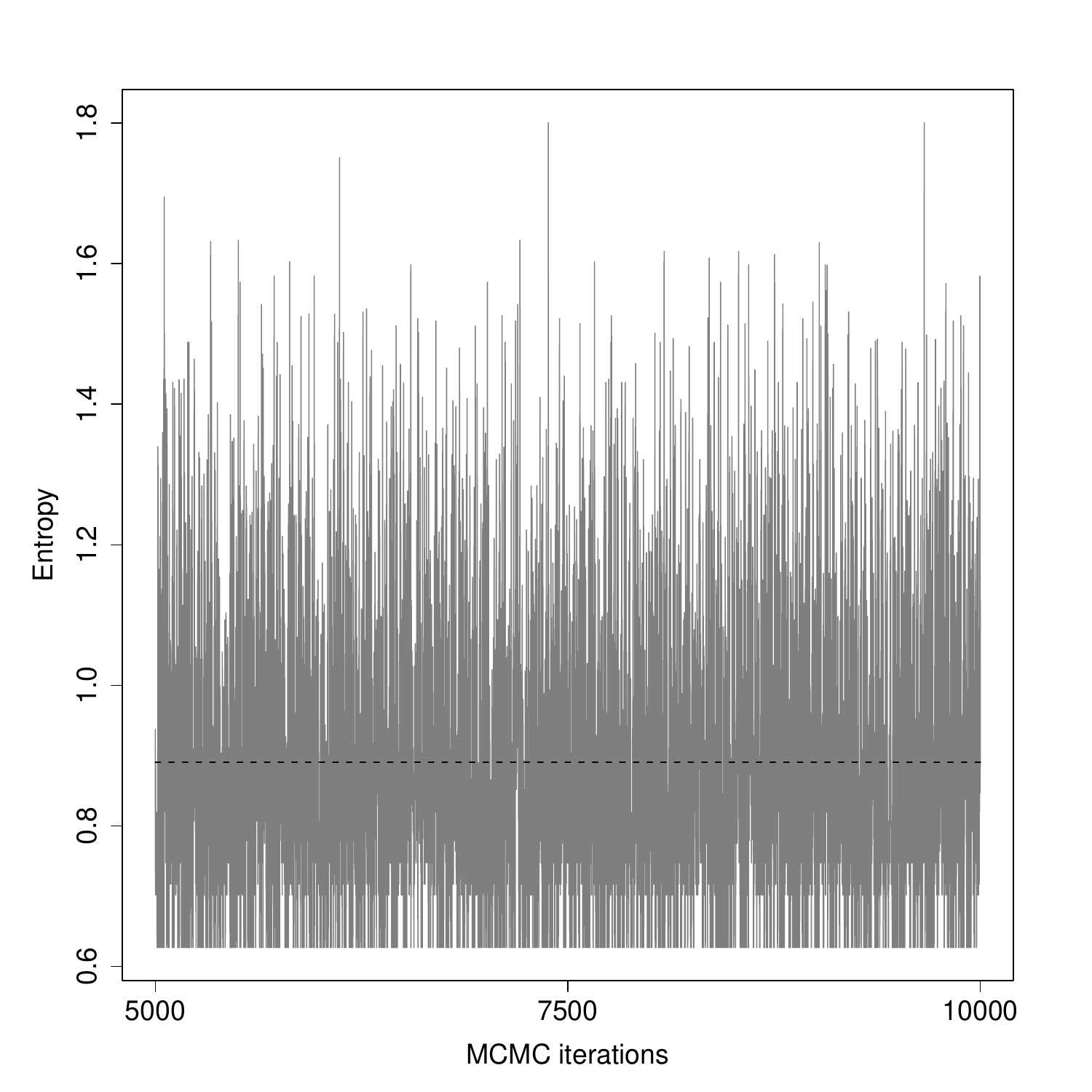}
 \caption{ trace plot of the entropy $H_t$ for a randomly selected replicate at a fixed time $t$. The dashed line indicates the mean entropy value. See Section \ref{sec:indep} for details.}\label{fig:entropy}
\end{figure}

Figure \ref{fig:entropy} displays the trace plot of the entropy $H_t$ of the partition at time $t$, based on a randomly selected replicate, where  
\begin{equation*}
H_t = \sum_{j=1}^{k_t} \frac{n_{jt}}{n}\log\left(\frac{n_{jt}}{n}\right),
\end{equation*}  
with $k_t$ denoting the number of clusters at time $t$, and $n_{jt}$ the number of units assigned to the $j$th cluster at that time.

\begin{figure}[h!]
\centering
 \includegraphics[clip,trim=0cm 0cm 0cm 0.15cm,width=0.65\textwidth]{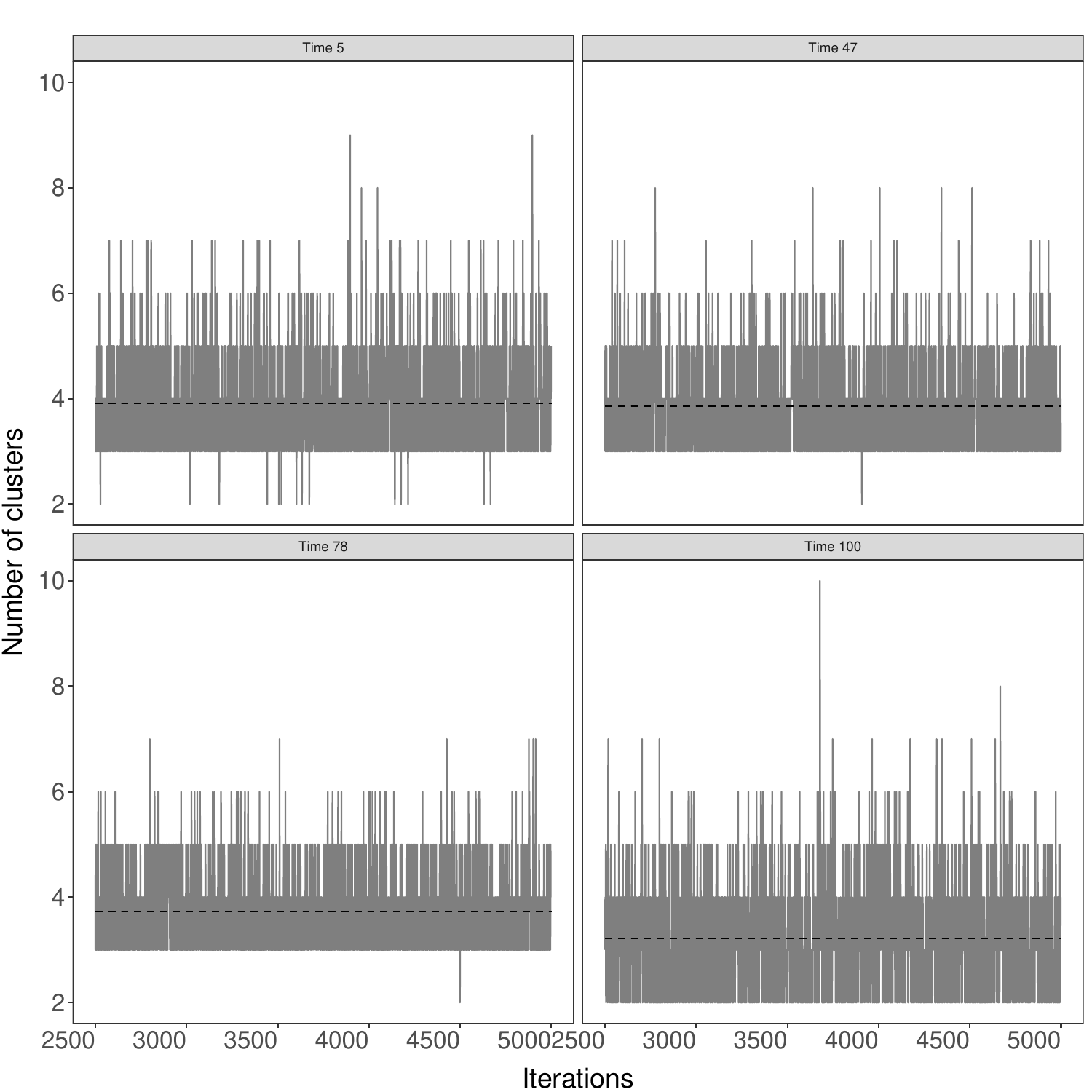}
 \caption{ trace plots of the number of cluster for a randomly selected replicate, for times $t = \{5,47,78,100\}$. See Section \ref{sec:indep} for details.}\label{fig:trace plots}
\end{figure}

\begin{figure}[t!]
    \centering
    \includegraphics[clip,trim=0cm 0cm 0cm 0.15cm,width=0.85\textwidth]{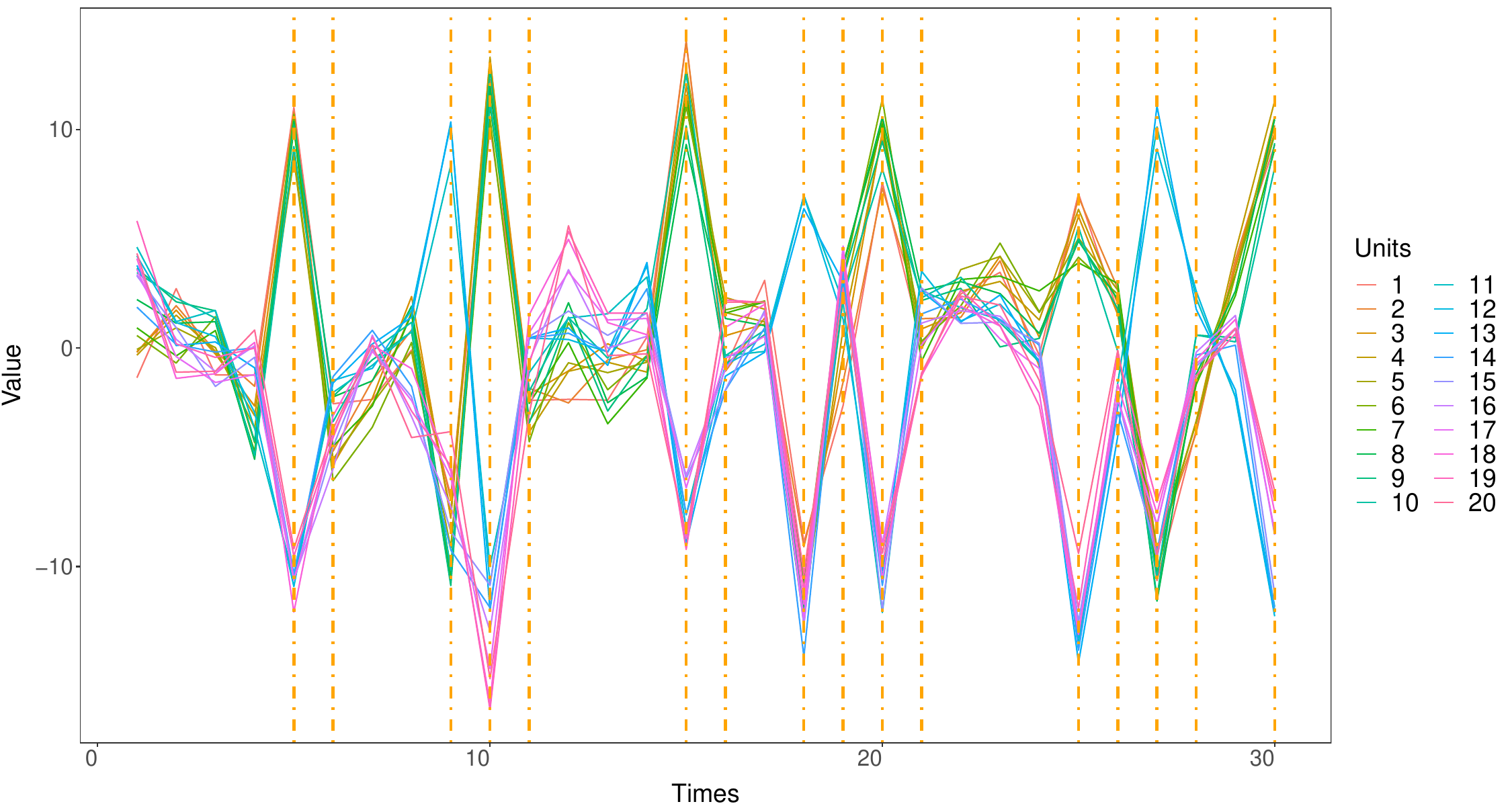}
    \caption{Example of simulated autoregressive dataset, with $n=20$ subjects and 30 time points. The orange vertical lines correspond to changepoints. The autoregressive coefficient is 0.9. See Section \ref{sec:ar1_data} for details.}
    \label{fig:ar1_res_cgp_lambda0.9}
\end{figure}

\clearpage
\section{Gesture Phase Segmentation}\label{app:gesture}
In this Section we represent additional plots referring to the application presented in Section \ref{sec:gesturedata}.

\begin{figure}[h!]
    \centering
    \includegraphics[clip,trim=0cm 0cm 0cm 0.15cm,width=0.85\textwidth]{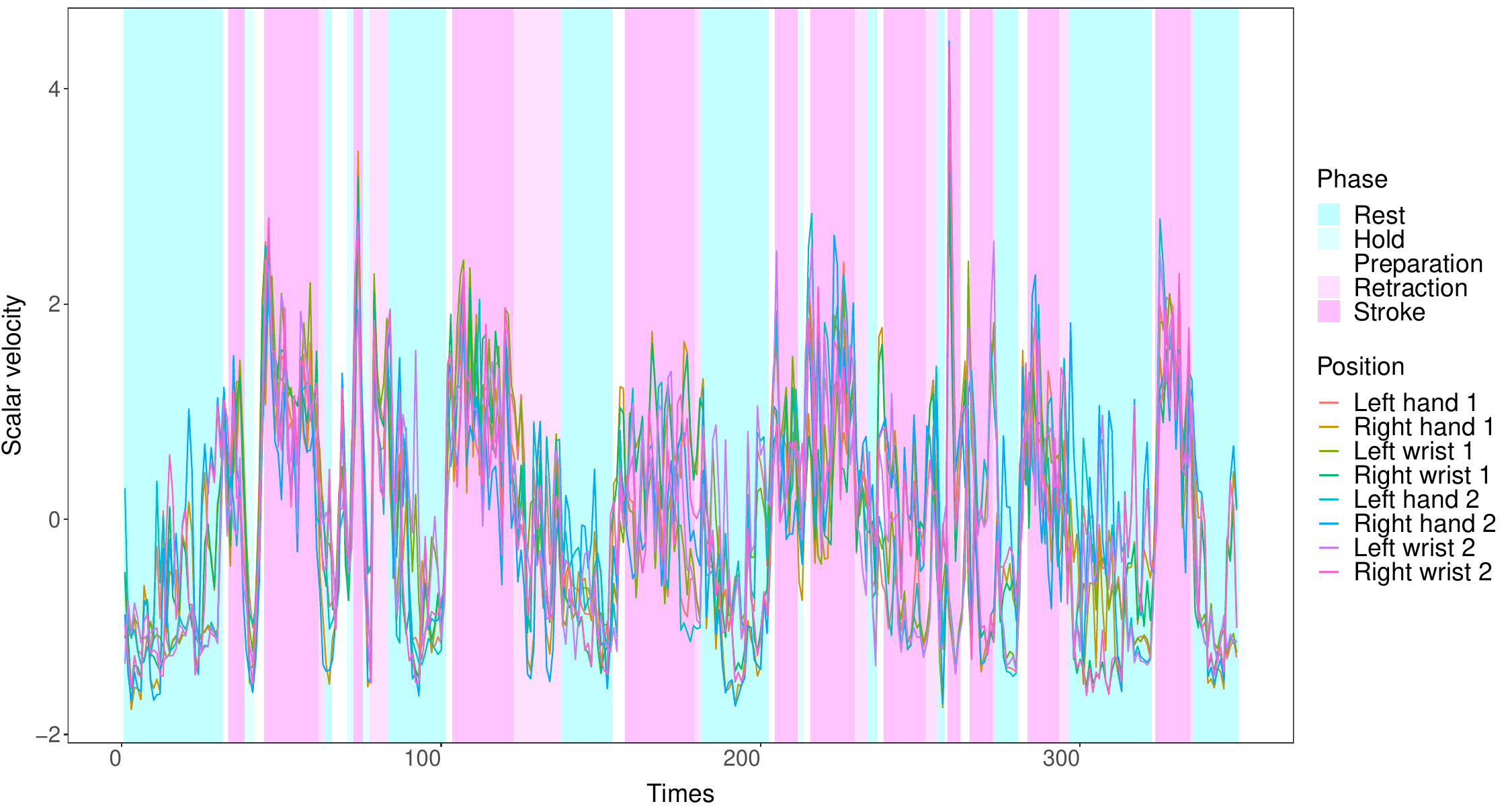}
    \caption{Human Gesture data with the phases of the video on the background, as estimated by \citet{madeo2013gesture}. See Section \ref{sec:gesturedata} for details.}
    \label{fig:gesture_both}
\end{figure}

\begin{figure}[ht]
    \centering
    \includegraphics[clip,trim=0cm 0cm 0cm 0.15cm,width=0.85\textwidth]{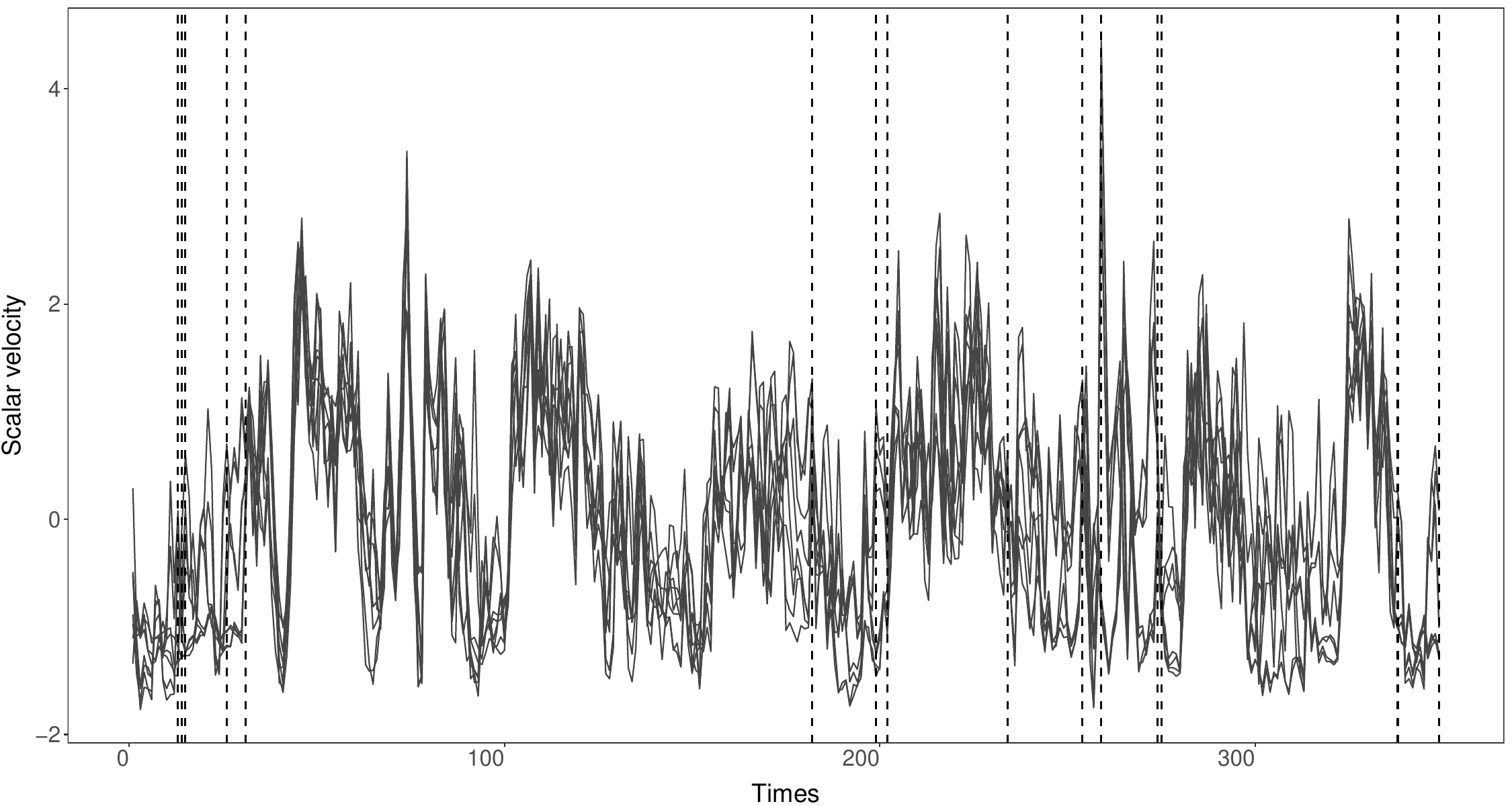}
    \caption{Human Gesture data.  Changepoints (dashed line) identified with the DRPM. 
    See Section \ref{sec:gesturedata} for details.}
    \label{fig:drpm_K2_2phases}
\end{figure}

\begin{figure}[ht]
    \centering
    \includegraphics[clip,trim=0cm 0cm 0cm 0.15cm,width=0.85\textwidth]{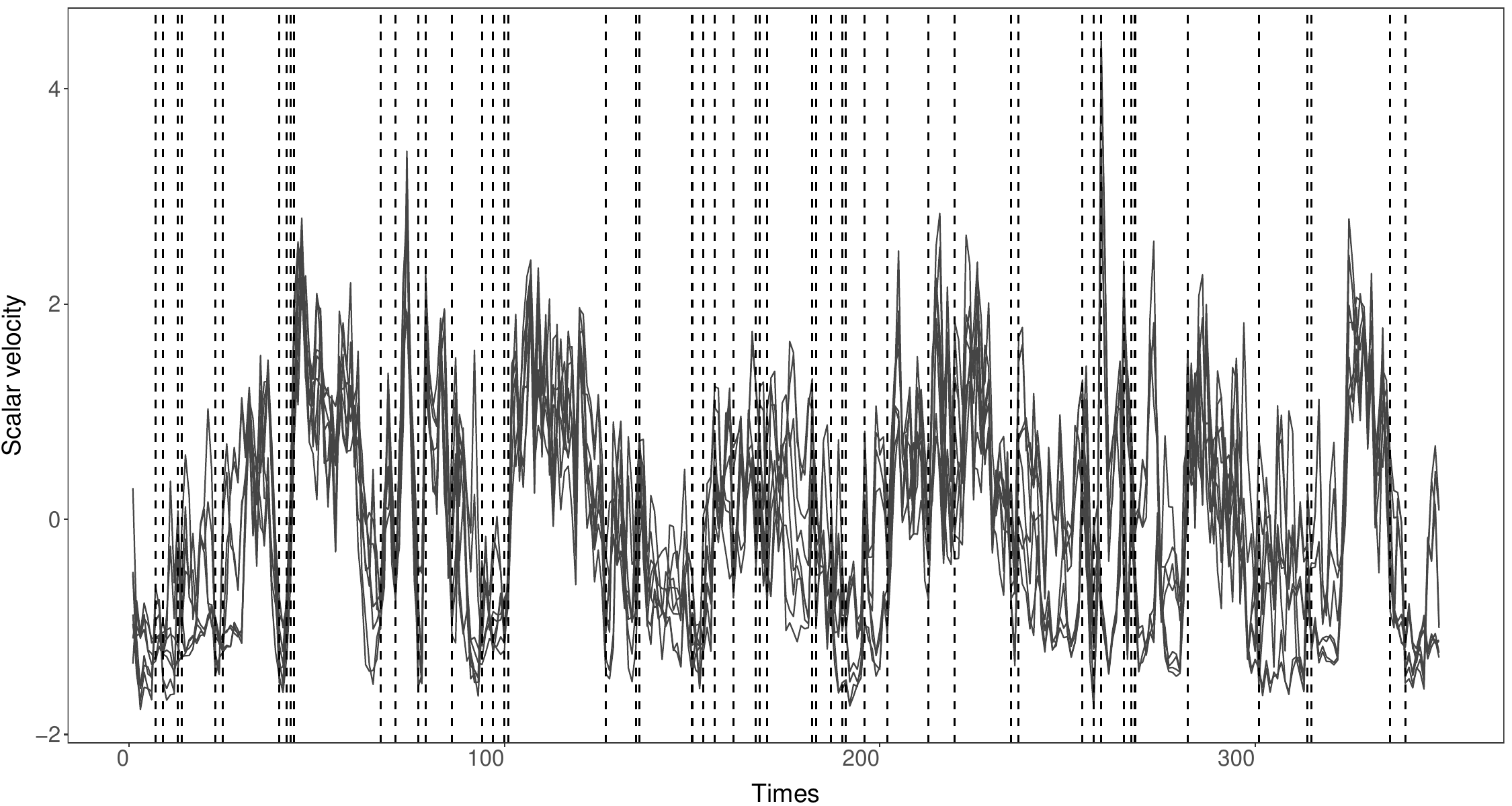}
    \caption{Human Gesture data.  Changepoints (dashed line) identified with the LDDP model. 
    See Section \ref{sec:gesturedata} for details.}
    \label{fig:lddp_K2_2phases}
\end{figure}
\begin{figure}[ht]
    \centering
    \includegraphics[clip,trim=0cm 0cm 0cm 0.15cm,width=0.85\textwidth]{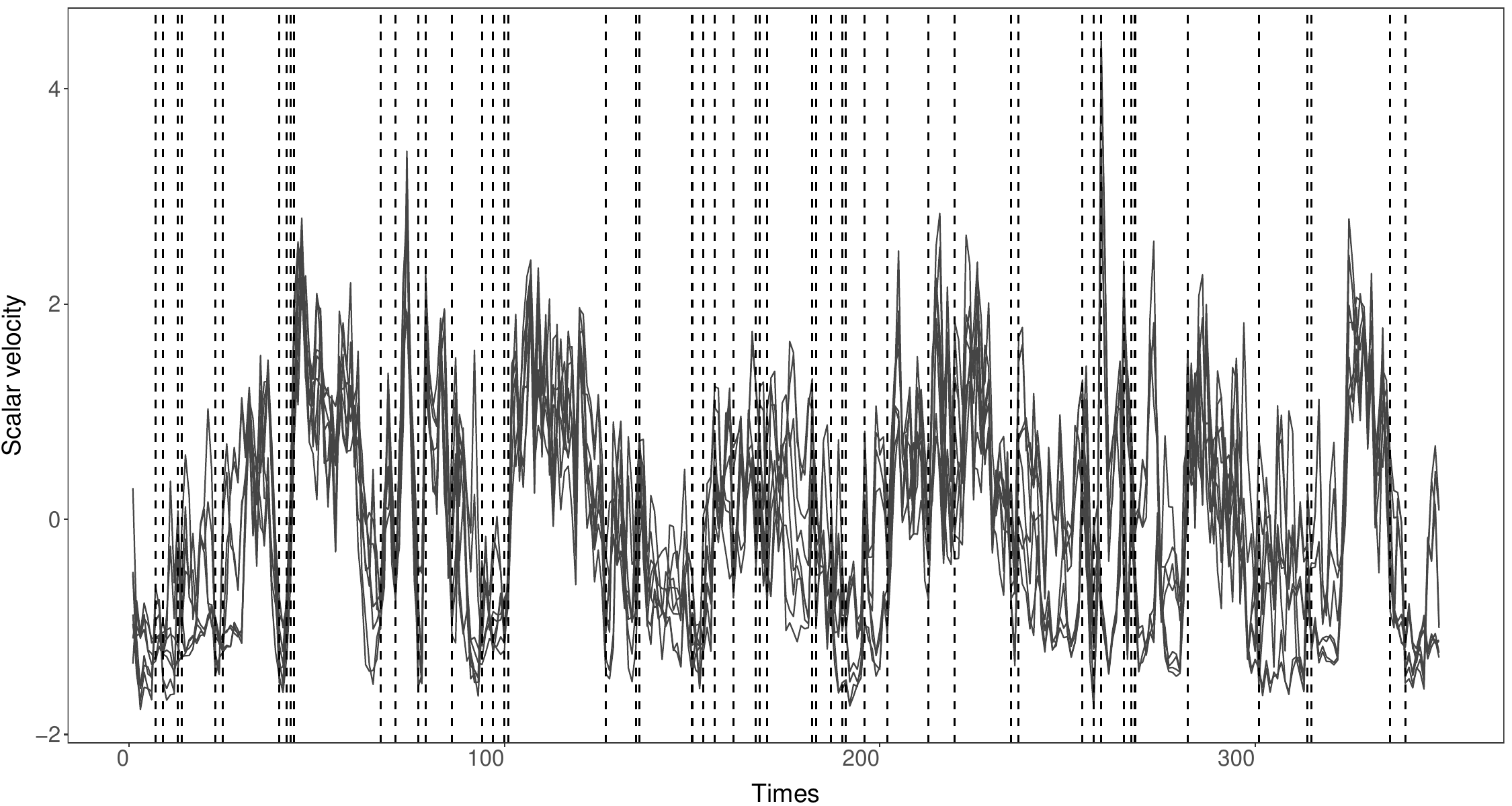}
    \caption{Human Gesture data.  Changepoints (dashed line) identified with the WDDP model. 
    See Section \ref{sec:gesturedata} for details.}
    \label{fig:wddp_K2_2phases}
\end{figure}

\begin{figure}[ht]
    \centering
    \includegraphics[clip,trim=0cm 0cm 0cm 0.15cm,width=0.85\textwidth]{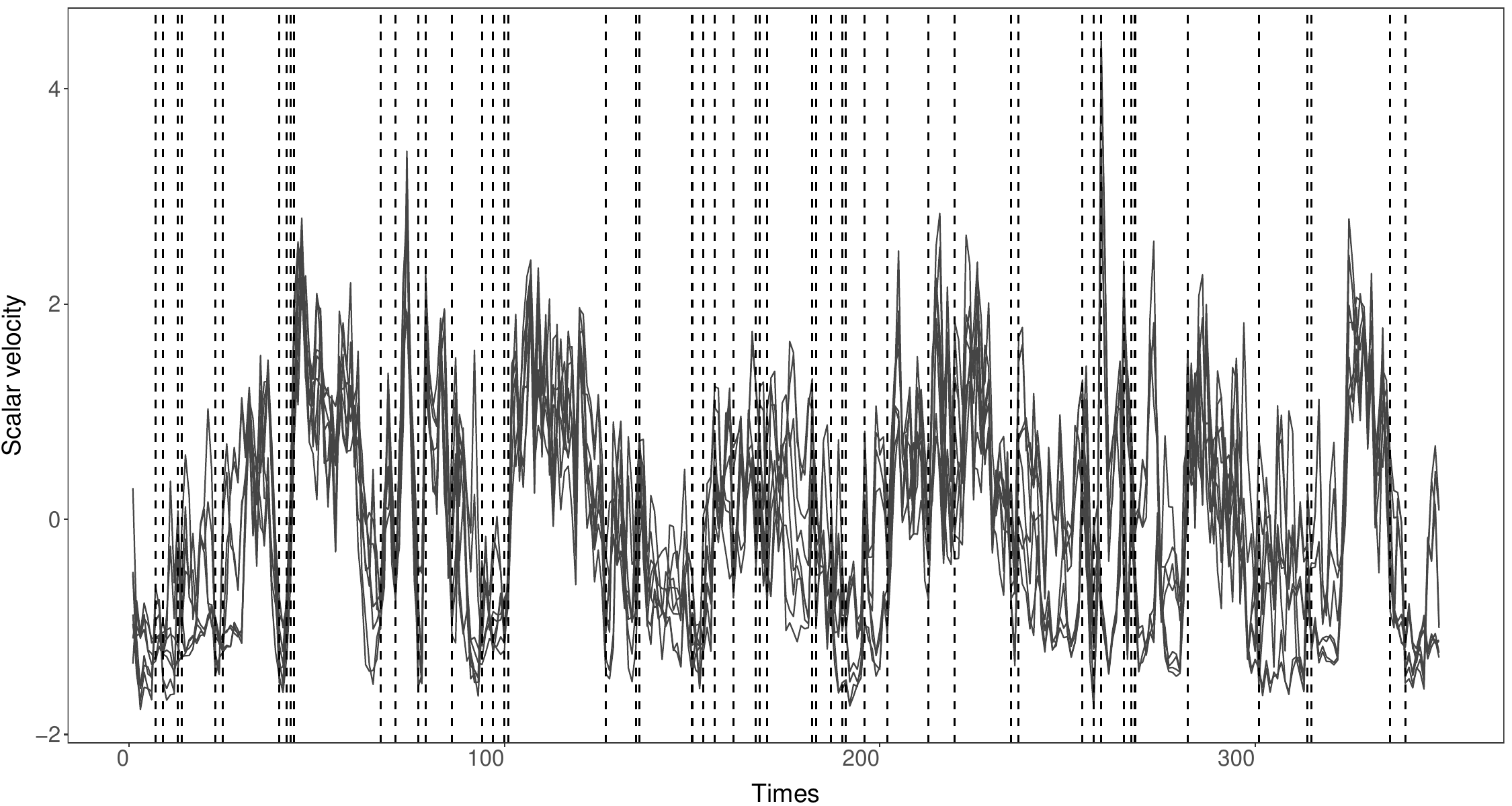}
    \caption{Human Gesture data.  Changepoints (dashed line) identified with the GMDDP model. 
    See Section \ref{sec:gesturedata} for details.}
    \label{fig:gmddp_K2_2phases}
\end{figure}

\end{document}